\documentclass[11pt]{article}
\pdfoutput=1
\usepackage{hyperref}
\usepackage{graphicx}
\usepackage{graphics}
\usepackage{dsfont}
\usepackage{epsfig}
\usepackage{amsmath,amssymb,amsthm,amscd}

\numberwithin{equation}{section}

\begin{document}
\begin{titlepage}
{}~ \hfill\vbox{ \hbox{} }\break

\rightline{USTC-ICTS-14-09}

\vskip 3 cm

\centerline{\Large
\bf
Topological Strings and Quantum Spectral Problems}
\vskip 0.5 cm

\renewcommand{\thefootnote}{\fnsymbol{footnote}}
\vskip 30pt \centerline{ {\large \rm Min-xin Huang\footnote{minxin@ustc.edu.cn}  and Xian-fu Wang\footnote{wangxf5@mail.ustc.edu.cn} } } \vskip .5cm \vskip 30pt

\begin{center}
{Interdisciplinary Center for Theoretical Study,  \\ \vskip 0.2 cm  Department of Modern Physics, University of Science and Technology of China,  
\\ \vskip 0.2 cm 96 Jinzhai Road,  Hefei, Anhui 230026, China}
\end{center}

\setcounter{footnote}{0}
\renewcommand{\thefootnote}{\arabic{footnote}}
\vskip 60pt
\begin{abstract}

We consider certain quantum spectral problems appearing in the study of local Calabi-Yau geometries. The quantum spectrum can be computed by the Bohr-Sommerfeld quantization condition for a period integral. For the case of small Planck constant,  the periods are computed perturbatively by deformation of the $\Omega$ background parameters in the Nekrasov-Shatashvili limit. We compare the calculations with the results from the standard perturbation theory for the quantum Hamiltonian. There have been proposals in the literature for the non-perturbative contributions based on singularity cancellation with the perturbative contributions. We compute the quantum spectrum numerically with some high precisions for many cases of Planck constant. We find that there are also some higher order non-singular non-perturbative contributions, which are not captured by the singularity cancellation mechanism. We fix the first few orders formulas of such corrections for some well known local Calabi-Yau models.

\end{abstract}

\end{titlepage}
\vfill \eject


\newpage

\baselineskip=16pt

\tableofcontents

\section{Introduction}

It is often fruitful to study the behavior of a theory at strong coupling, which may be related to another theory at weak coupling. Today we have many understandings of the non-perturbative effects in string theory, due to the studies of D-branes and string dualities in the middle 1990's.  However, a full non-perturbative formulation of superstring theory is still lacking. We may try to attack the problem in some simpler settings. Some important lessons were provided by the studies of non-critical string theory described by matrix models in early 1990's.  In these simpler models one can have better handle on the string perturbation series, and the studies of their large order behaviors often reveal the nature of non-perturbative effects. See e.g. \cite{DiFrancesco:1993, Marino:2012} for reviews on the subject.

Topological string theory has been very useful for counting holomorphic curves on Calabi-Yau spaces, and also has many other applications \cite{Hori}. Recently, there have been some research on the refined topological string theory. This is motivated by the $\Omega$ background, proposed for the purpose of calculating partition functions of Seiberg-Witten theories \cite{Nekrasov}, and is also applied for more general theories with quiver gauge groups \cite{Nekrasov:2013}. The refined topological string partition function on non-compact toric Calabi-Yau geometries can be computed by the A-model method of refined topological vertex \cite{IKV}, generalizing the earlier work of topological vertex \cite{AKMV:2003}. This has been related to the partition function of M-branes \cite{Haghighat:2013}. On the other hand, it can be also computed by B-model method using mirror symmetry \cite{HK:2010, KW}, which generalizes the holomorphic anomaly equation \cite{BCOV} and gap boundary conditions \cite{HK:2006} in the conventional unrefined case. Furthermore, the B-model approach can also work for certain non-toric del Pezzo Calabi-Yau geometries \cite{Huang:2013}.

There are two small expansion parameters $\epsilon_1, \epsilon_2$ in refined topological string theory, the conventional unrefined case corresponds to $\epsilon_1+\epsilon_2=0$. The worldsheet formulation with two expansion parameters is not so clear as the unrefined case where the expansion parameter counts the worldsheet genus. Some attempts are made in \cite{Nakayama:2011, Huang:2011, Antoniadis1} for clarifying the issue. The mathematical definition in terms of stable pair invariants is provided in \cite{Choi:2012}. See also \cite{Hellerman:2012, Antoniadis2} for the construction of the $\Omega$ background from superstring theory compactifications.

Another interesting limit is to set one of the $\epsilon_{1,2}$ to zero, known as the Nekrasov-Shatashvili limit \cite{Nekrasov:2009}, with deep connections to quantum integrable systems. The gauge theory and topological string partition function in this limit can be computed by deformed periods \cite{MM, Poghossian, Aganagic:2011, Huang:2012}. The Calabi-Yau geometry is related to a quantum mechanical Hamiltonian, and the deformed period is the phase volume which can be used to compute the energy spectrum of the Hamiltonian by the Bohr-Sommerfeld (BS) quantization condition.

The main purpose of this paper is to study non-perturbative effects in refined topological string theory. Some proposals have been made recently in \cite{Lockhart, Hatsuda:2013}. The non-perturbative sectors may also have holomorphic anomaly equation similarly as the perturbative sector \cite{Santamaria:2013}. Topological string is also an ideal place for the studies since the A-model amplitudes are exact in string coupling constant, essentially summing up all genus contributions, although at a finite degree of cohomology class. The proposal of \cite{Hatsuda:2013} is based on the relation between the local $\mathbb{P}^1\times \mathbb{P}^1$ Calabi-Yau model with the ABJM  (Aharony-Bergman-Jafferis-Maldacena) matrix models \cite{ABJM}. The ABJM theory is a 3d Chern-Simon theory dual to M-theory on $AdS_4\times S^7/\mathbb{Z}_k$, and its partition function on $S^3$ localizes to a matrix model \cite{Kapustin}.  Certain non-perturbative contributions are proposed to cancel the singularity encountered in the calculations of the partition functions of the ABJM matrix model \cite{Calvo, HMO, HMO1, HMO2, Hatsuda:2013, Honda:2014}, known as the Hatsuda-Moriyama-Okuyama (HMO) mechanism. The Wilson loops in the theory have been also studied extensively in the literature, see e.g. the recent work \cite{Hirano:2014}.

Since the quantum Hamiltonian related to the  local Calabi-Yau geometry is well defined for any Planck constant and the energy spectrum can be calculated numerically,  it is an ideal testing ground for the non-perturbative contributions in refined topological string theory \cite{Kallen:2013, Grassi:2014}. In \cite{Kallen:2013}, Kallen and Marino find that the perturbative B-period, i.e the quantum phase volume, is singular for infinitely many values of Planck constant $\hbar$, and they introduce the novel idea that the singularities would be cured by non-perturbative instanton contributions, which we shall call the Kallen-Marino (KM) singularity cancellation mechanism. The authors stress that this is not a consequence of the usual story of non-perturbative/perturbative completion, since the divergence of the perturbative series is not due to the factorial growth of its coefficients. Their study is based on the 
$\mathbb{P}^1\times \mathbb{P}^1$ model dual to the ABJM matrix model, but they also propose to consider the quantum spectral problems for general local Calabi-Yau spaces such as the local $\mathbb{P}^2$ geometry  at the end of the paper.

In this paper, we shall push the idea to some fruitions. We consider some well-known local Calabi-Yau geometries, namely the local $\mathbb{P}^2$, $\mathbb{P}^1\times \mathbb{P}^1$ and $\mathbb{F}_1$ models. First we study the perturbative expansion of the spectrum for small $\hbar$ and use two methods for the computations. Then we consider non-perturbative effects, and find that the requirement of Kallen-Marino singularity cancellation largely fixes the singular part of the non-perturbative contributions to the quantum phase volume. The remaining ambiguity can be fixed by checking with the numerical calculations of the quantum spectrum.

However, the Kallen-Marino singularity cancellation mechanism is not the whole story. We further consider some samples of specific values of the Planck constant and test the proposal for non-perturbative contributions with numerical calculations. We discover that there are certain higher order non-singular corrections in the non-perturbative contributions, which do not affect the singularity cancellation with perturbative contributions. For the case of the local $\mathbb{P}^2$ model, their effects first show up at the 3rd sub-leading order in the large energy expansion, and can only be discovered by some  high precision numerical calculations. With the results of the calculations for the samples of the Planck constant, we can guess the exact formulas for the first few orders of such corrections.

We should note that our formulation of the quantum Hamiltonian for the $\mathbb{P}^1\times \mathbb{P}^1$ model is quite different from the one dual to the ABJM matrix model in \cite{Kallen:2013}. In the ABJM formulation, the Hamiltonian comes from an integral equation determining the spectrum with a Hilbert-Schmid kernel. There are well-known existence theorems in the elementary theory of integral equations that the quantum spectral problem is well defined. On the other hand, our formulation of the Hamiltonian is more natural for topological string theory since it can be applied to general local toric Calabi-Yau geometries. Although we are not aware of a mathematical proof that the spectral problem for our Hamiltonian is well defined, we can still calculate the discrete spectrum numerically in an orthonormal basis of wave functions for any Planck constant. As a result we believe our formulation is also consistent.  At the classical level, the spectral curves of the two formulations are related by a coordinate transformation \cite{Hatsuda:2013, Kallen:2013}. However, at the quantum level, the corresponding spectra are quite different and we are not aware of a simple transformation that relates them. As such our Hamiltonian for the $\mathbb{P}^1\times \mathbb{P}^1$ model may not be much relevant for the studies of the ABJM matrix model. It would be still interesting to see whether the higher order non-perturbative contributions we find are also present for the ABJM formulation of the  $\mathbb{P}^1\times \mathbb{P}^1$ Hamiltonian.  

The organization of the paper is the followings. In Section \ref{P2section}, we consider in details our main example, the local $\mathbb{P}^2$ model. Our method can be straightforwardly applied to other  local Calabi-Yau models, such as the ones from anti-canonical bundle over del Pezzo surfaces, constructed by blowing up points on  the $\mathbb{P}^2$ geometry. One can also consider the Hirzebruch surfaces, which are $\mathbb{P}^1$ bundles over $\mathbb{P}^1$. The differential operators for the deformed periods are studied in \cite{Huang:2012, Huang:2014}.  In Sections \ref{P1P1section}, \ref{F1section}  we study two such examples, namely the local $\mathbb{P}^1\times \mathbb{P}^1$ and $\mathbb{F}_1$ models. Here the local $\mathbb{P}^1\times \mathbb{P}^1$ model can be regarded as in the class of both del Pezzo surfaces and Hirzebruch surfaces. We present the results with less details since the method is similar to the main example. Our main result are the non-perturbative formulas (\ref{mainresult}, \ref{mainresultP1P1}, \ref{mainresultF1}) for the three examples.

\section{The local $\mathbb{P}^2$ model} \label{P2section}

Our main example is  the local $\mathbb{P}^2$ model, well-known in the mirror symmetry literature. The geometry is described by the classical curve on $(x,p)$ plane
\begin{eqnarray} \label{curve2.1}
e^x + e^p + z e^{-x} e^{-p} =1,
\end{eqnarray}
where $z$ is the complex structure modulus parameter of the geometry.

The Hamiltonian operator is derived from the curve (\ref{curve2.1}) by the following rescaling and shifts
\begin{eqnarray} \label{scaling2.2}
z\rightarrow e^{-3H}, ~~~ x\rightarrow x - H, ~~~ p\rightarrow p - H
\end{eqnarray}
Furthermore, we promote the $x,p$ to the quantum position and momentum operators, satisfying the canonical commutation relation $[\hat{x}, \hat{p} ] = i \hbar$. We then find the one-dimensional quantum mechanical Hamiltonian
\begin{eqnarray} \label{Hamiltonian2.3}
\hat{H} = \log( e^{\hat{x}} + e^{\hat{p}} + e^{ -\hat{x}-\hat{p}} ).
\end{eqnarray}
We note that the Hermitian condition uniquely determines the ordering of the last term. For example, the following different orderings are actually the same
\begin{eqnarray}
 e^{ -\frac{\hat{x}}{2}} e^{- \hat{p}} e^{ -\frac{\hat{x}}{2}} =  e^{ -\frac{\hat{p}}{2}} e^{- \hat{x}} e^{ -\frac{\hat{p}}{2}}
 =e^{-\frac{i\hbar}{2}} e^{- \hat{x}}e^{-\hat{p}}=  e^{- \hat{x}-\hat{p}} ,
\end{eqnarray}
due to the Baker-Campbell-Hausdorff formula.

In the scaling (\ref{scaling2.2}) we can also keep the $z$ parameter by using $z\rightarrow z e^{-3H}$ instead. The studies of the resulting Hamiltonian are related to the one in (\ref{Hamiltonian2.3}) by a simple transformation. For simplicity we will not keep this parameter.

Comparing to the  local $\mathbb{P}^1\times \mathbb{P}^1$ model in \cite{Kallen:2013}, the exponentiated Hamiltonian from (\ref{Hamiltonian2.3}) can not be written as a product of several factors. The quantum Hamiltonian should have a discrete spectrum bounded below for any real value of Planck constant $\hbar$. The quantum spectral problem is difficult to solve, since the Schrodinger equation involves infinitely many higher derivatives in the position space. We should use the old Bohr-Sommerfeld quantization method
\begin{eqnarray} \label{BS2.6}
\textrm{vol}(E) = 2\pi \hbar (n+\frac{1}{2}),
\end{eqnarray}
where the volume in phase space is defined by period integral $\textrm{vol}(E)  \equiv \oint p(x) dx$. This approach is proposed by Nekrasov and Shatashvili in the context of $\mathcal{N}=2$ supersymmetric gauge theory \cite{Nekrasov:2009}. In the classical limit, the period integral is simply the B-period of the local Calabi-Yau geometry. In the quantum theory, we shall consider the refined topological string theory and take the Nekrasov-Shatashvili limit where one of the $\epsilon_{1,2}$ parameters of the $\Omega$ background is set to zero, and the other is identified with the Planck constant $\hbar$. The volume $\textrm{vol}(E)$ is then computed by the deformed B-period in the Nekrasov-Shatashvili limit.

\subsection{Classical ground state energy} \label{sectionclassical}
In the small $\hbar$ limit, we can expand the energy spectrum as
\begin{eqnarray} \label{En2.6}
E^{(n)} = \sum_{k=0}^\infty E^{(n)}_k \hbar^k  .
\end{eqnarray}
The classical  ground state energy is the minimum of the classical potential, and should be the same for all quantum levels. We denote the classical ground state energy as $E_0 = E^{(n)}_0 $ for any quantum level $n$.

It is easy to compute $E_0$ by taking the classical limit $\hbar\rightarrow 0$. We can work in the position space and the momentum operator $\hat{p} = -i\hbar \partial_x \rightarrow 0$ in this limit. We find
\begin{eqnarray}
\hat{H} \rightarrow \log(e^{x} + e^{-x} +1  ) \geq \log(3),
\end{eqnarray}
where the equality is saturated at $x=0$. So the classical minimum energy is $E_0=\log(3)$.

To illustrate the idea of computing the quantum spectrum by the Bohr-Sommerfeld quantization method, we first apply it in the simple case of the classical limit. We denote the classical volume $\textrm{vol}_0(E)$, and the Bohr-Sommerfeld quantization condition in the classical limit $\hbar=0$ is simply
\begin{eqnarray} \label{classical2.8}
\textrm{vol}_0(E_0) =0 .
\end{eqnarray}
In the followings we should compute the classical volume $\textrm{vol}_0(E)$, and reproduce the  classical minimum energy $E_0=\log(3)$ from the above equation (\ref{classical2.8}).

The topological string on the local $\mathbb{P}^2$ model and its modularity were studied in details in \cite{ABK, HKR}. The periods are determined by the well-known Picard-Fuchs differential equation
\begin{eqnarray} \label{PF2.9}
[ \Theta_z^3 - 3z(3\Theta_z+2)( 3\Theta_z+1) \Theta_z ] w(z) =0,
\end{eqnarray}
where the differential operator is defined as $\Theta_z := z\partial_z$. There are three linearly independent solutions to the differential equation, and can be obtained by the following Frobenius method. Define the infinite series
\begin{eqnarray}
w(z,s) = \sum_{n=0}^{\infty} \frac{ (-1)^n z^{s+n}} {\Gamma(-3(n+s)+1) \Gamma^3(n+s+1)} ,
\end{eqnarray}
then the solutions to the differential equation (\ref{PF2.9}) can be obtained by $w_k(z) = \frac{d^k}{d^k s} w(z,s) |_{s=0}$. Taking $k=0,1,2$, we find the three  linearly independent  series solutions
\begin{eqnarray} \label{periods2.11}
w_0 =1, ~~~ w_1(z) =\log(z) + \sigma_1(z), ~~~~ w_2(z) =(\log z)^2 + 2\sigma_1 (\log z) + \sigma_2(z),
\end{eqnarray}
where $w_1(z)$ and $w_2(z)$ are the logarithmic and double-logarithmic solutions, usually known as A-period and B-period of the geometry, and the power series are defined by the Digamma function $\psi(x) =\frac{\Gamma^{\prime}(x)}{\Gamma(x)}$ as
\begin{eqnarray}
\sigma_1(z) = \sum_{n=1}^{\infty} 3 z^n \frac{(3n-1)!}{n!^3}, ~~~ \sigma_2(z) =  \sum_{n=1}^{\infty} 18 z^n \frac{(3n-1)!}{n!^3} [\psi(3n) -  \psi(n+1) ].
\end{eqnarray}
After substituting the parameter $z= e^{-3E}$, we see that in large $E$ limit, the logarithmic terms in the periods provide finite contributions, while the power series $\sigma_{1,2}(z) $ give exponentially small corrections.

We can also solve the equation (\ref{PF2.9}) near the conifold point $z\sim \frac{1}{27}$. Denoting the small parameter $z_c=1/27 - z$, the three linearly independent solutions are
\begin{eqnarray}
&& t_0=1,~~~ t_1(z)= z_c + \frac{33 z_c^2}{2} + 327z_c^3  +\frac{28167 z_c^4}{4}+\mathcal{O}(z_c^5)  , \nonumber \\
&& t_2(z) = t_1(z) \log(z_c) +\frac{63 z_c^2}{4} + \frac{877 z_c^3}{2} +\frac{176015 z_c^4}{16} + \mathcal{O}(z_c^5).  \label{conifold2.13}
\end{eqnarray}
We only need to consider the case of positive Planck constant $\hbar$, since the quantum Hamiltonian (\ref{Hamiltonian2.3}) is invariant under the exchange of position $\hat{x}$ and momentum $\hat{p}$, which changes the sign of $\hbar$. We will see that  the quantum energy $E\geq E_0 =\log(3)$ for $\hbar\geq 0$, so $z=e^{-3E}\leq \frac{1}{27}$. We have used the coordinate $z_c=1/27 - z$ so that $z_c\geq 0$ and the logarithmic cut $ \log(z_c) $ in $t_2$ is real. The three periods $t_0, t_1, t_2$ are linear combinations of $w_0,w_1, w_2$ in (\ref{periods2.11}) when one analytically continue from $z\sim0$ to $z\sim \frac{1}{27}$.

It turns out that the classical volume is not exactly the B-period $w_2$, but also contains a constant from the first period $w_0$, as shown for the local $\mathbb{P}^1\times\mathbb{P}^1$ model in \cite{Marino:2011}. In order to determine the correct constant, we shall follow the method similarly as \cite{Marino:2011}, and compute the classical volume $\textrm{vol}_0(E)$ in the large energy $E$ limit, neglecting the exponentially small corrections.

We can solve for the momentum from the Hamiltonian (\ref{Hamiltonian2.3}) at energy $E$ in the classical limit
\begin{eqnarray}
 p_{\pm} = \log[\frac{e^E-e^x \pm \sqrt{(e^E -e^x)^2-4e^{-x}  }}{2}].
 \end{eqnarray}
These two solutions provide a bounded area in the real $ (x,p) $ plane and define the classical phase volume, or more precisely the phase area
\begin{eqnarray} \label{volume2.14}
\textrm{vol}_0(E) = \int_{e^x+e^p +e^{-x-p} \leq e^E} dx dp = \int_{a}^{ b} (p_+(x) - p_-(x ) ) d x,
\end{eqnarray}
where the range of the definite integral $a,b$ are the two roots of the equation from the square root term $(e^E -e^x)^2-4e^{-x} =0$, so that $p_+(x) = p_-(x)$ at $x =a,b$, and satisfying  $(e^E -e^x)^2-4e^{-x}>0$ for $a<x<b$.

It is clear that for the classical ground state energy $E_0=\log(3)$, the phase space has only one point $(x,p)=(0,0)$ and therefore the volume vanishes $\textrm{vol}_0(E_0)=0$. We wish to compute the classical volume $\textrm{vol}_0(E)$ for arbitrary $E\geq E_0$.

\begin{figure}[h!]
\begin{center}
\includegraphics[angle=0,width=0.6\textwidth]{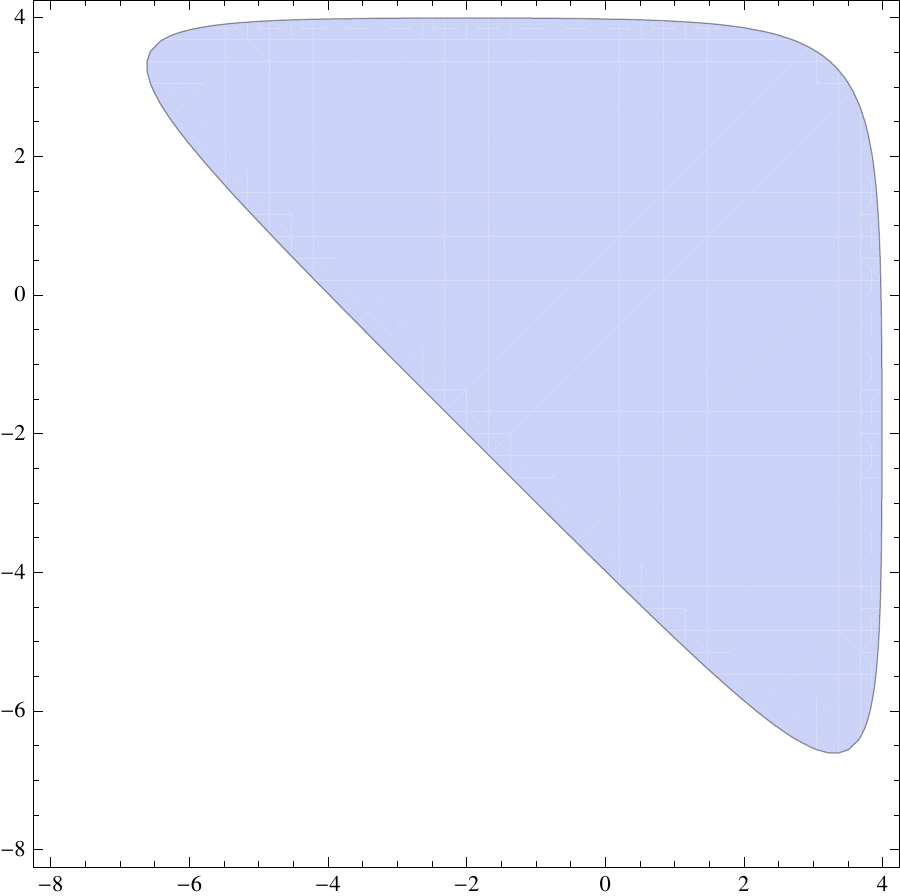}
\begin{quote}
\caption{ The phase space in the real $(x,p)$ place, parametrized by the equation  $e^x+e^p +e^{-x-p} \leq e^E$, for the example of $E=4$.
\vspace{-1.2cm}} \label{figurephasevolume}
\end{quote}
\end{center}
\end{figure}

The integral is quite complicated to do exactly, but the computation becomes much simpler if we can neglect exponentially small corrections in large $E$. The integration range is then
\begin{eqnarray}
a = -2E +\log(4) +\mathcal{O}(e^{-E}), ~~~  b = E + \mathcal{O}(e^{-E}).
\end{eqnarray}
We can see that in the large $E$ limit, the phase space asymptotes to roughly the shape of a triangle, depicted in Figure \ref{figurephasevolume}.

We compute the phase volume by plugging the formulae for $p_{\pm}$, and we find
\begin{eqnarray}
\textrm{vol}_0(E) &=& \int_{-2E +\log(4)}^E \{ 2E +x + 2\log[\frac{1-e^{x-E} + \sqrt{(1 -e^{x-E})^2-4e^{-x-2E}  }}{2}] \} dx \nonumber \\
&=& \frac{9E^2}{2} - 2\log^2(2)+ 2\int_{-2E +\log(4)}^E \log[\frac{1-e^{x-E} + \sqrt{(1 -e^{x-E})^2-4e^{-x-2E}  }}{2}] dx.
\nonumber
\end{eqnarray}

Suppose $ x_0\in (-2E+\log 4, E) $ is a generic value in the integration range, with $x_0+2E\sim E-x_0 \sim E $ in the large $E$ limit. We divide the definite integral into two parts, and neglect exponentially small corrections
\begin{eqnarray}
\textrm{vol}_0(E) =  \frac{9E^2}{2} - 2\log^2(2) + 2\int_{x_0 }^E \log(1-e^{x-E} ) dx + 2\int_{-2E +\log(4)}^{x_0} \log[\frac{1 + \sqrt{1 -4e^{-x-2E}  }}{2}] dx . \nonumber
\end{eqnarray}
The first integral is simple to compute
\begin{eqnarray}\label{firstint}
\int_{x_0 }^E \log(1-e^{x-E} ) dx = - \sum_{k=1}^{\infty} \int_{x_0 }^E \frac{e^{k(x-E)}}{k} dx  = -\sum_{k=1}^{\infty} \frac{1}{k^2} =  - \frac{\pi^2}{6}  .
\end{eqnarray}
For  the second integral, we use the following indefinite integral with the polylogarithmic function
\begin{eqnarray}
\int^x \log[ \frac{1+\sqrt{1-e^{c-x }} }{2} ] dx &=&  \textrm{Li}_2 (  \frac{1- \sqrt{1-e^{c-x }} }{2} ) +c\cdot \textrm{arctanh} ( \sqrt{1-e^{c-x }})-\frac{1}{4} (c^2 + 2cx)  \nonumber \\
&&  - c \log[ \frac{1+\sqrt{1-e^{c-x }} }{2} ] -\frac{1}{2} \log^2 [ \frac{1+\sqrt{1-e^{c-x }} }{2} ]  .
\end{eqnarray}
The definite integral can be evaluated by plugging in the integration range, we find that the result is also independent of the specific value of $x_0$
\begin{eqnarray}
\int_{-2E +\log(4)}^{x_0} \log[\frac{1 + \sqrt{1 -4e^{-x-2E}  }}{2}] dx = -\textrm{Li}_2(\frac{1}{2}) +\frac{\log ^2(2)}{2}
= -\frac{\pi^2}{12} + \log ^2(2) .
\end{eqnarray}

Summarizing the results of the calculations, we find that
\begin{eqnarray}
\textrm{vol}_0(E) = \frac{9E^2 -\pi^2}{2} +\mathcal{O}(e^{-E}) .
\end{eqnarray}
So we see that the correct combination of periods in (\ref{periods2.11}) for the phase volume should be
\begin{eqnarray}
  \textrm{vol}_0(E) = \frac{1}{2} (w_2 - \pi^2)
 \end{eqnarray}
Including the full series in the period $w_2$ and replacing $z=  e^{-3E}$, we recover the full exponentially small corrections $\mathcal{O}(e^{-E})$ in  the classical volume
\begin{eqnarray} \label{classical2.22}
\textrm{vol}_0(E) =  \frac{9E^2 -\pi^2}{2}    +
 9\sum_{n=1}^{\infty} e^{-3nE}  \frac{(3n-1)!}{n!^3} [\psi(3n) -  \psi(n+1) -E]   .
 \end{eqnarray}
We can check numerically that the equation for the classical minimum energy $\textrm{vol}_0(E_0)=0$  is indeed an  identity for $E_0=\log(3)$. Of course, we can derive the classical minimum energy without the seemingly complicated computation of the phase volume. The Bohr-Sommerfeld quantization method would become essential later when we consider quantum and non-perturbative corrections when the Planck constant $\hbar$ is non-zero.

\subsection{Quantum perturbative contributions} \label{sectionper}
We consider the corrections to phase volume and energy eigenvalues that are powers of $\hbar$ in the small $\hbar$ expansion. From previous calculations of the deformed periods in local Calabi-Yau spaces, in e.g. \cite{Aganagic:2011, Huang:2012}, we expect the expansion of the phase volume has only even powers of $\hbar$. The energy spectrum, on the other hand, has corrections for integer powers of $\hbar$.  We denote the expansions as
\begin{eqnarray}
\textrm{vol}_p(E) = \sum_{k=0}^{\infty} \textrm{vol}_k(E) \hbar^{2k},~~~ E = \sum_{k=0}^{\infty} E_k \hbar^{k},
\end{eqnarray}
where the subscript $p$ denotes perturbative contributions. We can expand the quantum volume for small $\hbar$, and the first few terms are
\begin{eqnarray}
\textrm{vol}_p(E) &=& \textrm{vol}_0(E_0) + E_1 \textrm{vol}_0^{\prime} (E_0) \hbar
+ [ \textrm{vol}_1(E_0) + E_2  \textrm{vol}_0^{\prime} (E_0) + \frac{1}{2} E_1^2 \textrm{vol}_0^{\prime\prime} (E_0)]\hbar^2  \nonumber \\
&& + [ E_1 \textrm{vol}_1^{\prime}(E_0) + E_3 \textrm{vol}_0^{\prime} (E_0) +E_1E_2 \textrm{vol}_0^{\prime\prime} (E_0) +\frac{1}{6} E_1^3 \textrm{vol}_0^{\prime\prime\prime} (E_0) ]\hbar^3 +\mathcal{O}(\hbar^4)  \nonumber
\end{eqnarray}
We can use the Bohr-Sommerfeld equation (\ref{BS2.6}) to compute the perturbative corrections to energy spectrum recursively, if we know the values of the quantum volumes $\textrm{vol}_k(E)$ and their derivatives at the classical minimum energy $E_0=\log(3)$.

The first order corrections to spectrum $E^{(n)}_1$ depend only on the classical phase volume
\begin{eqnarray}  \label{E12.24}
E^{(n)}_1 = \frac{ (2n+1)\pi }{ \textrm{vol}_0^{\prime}(E_0)}.
\end{eqnarray}

We can check this formula directly from the Hamiltonian (\ref{Hamiltonian2.3}). The canonical commutation relation $[\hat{x},\hat{p}]=i\hbar $ implies that the contributions of the operators $\hat{x},\hat{p}$ are of order $\sqrt{\hbar}$ in the small $\hbar$ limit. In order to calculate the corrections up to order $\hbar$, we can expand the Hamiltonian
\begin{eqnarray} \label{expan2.25}
e^{\hat{H}} &=& 3 + \hat{x}^2 + \hat{p}^2 +\hat{x}\hat{p} -\frac{i\hbar}{2} + \mathcal{O}(\hbar^\frac{3}{2})  \\ \nonumber
&=& 3+ (\hat{x} +\frac{\hat{p}}{2})^2 +\frac{3}{4} \hat{p}^2 + \mathcal{O}(\hbar^\frac{3}{2}) .
\end{eqnarray}
We can redefine $\hat{x}^{\prime} = \hat{x} +\frac{\hat{p}}{2} $, which also satisfy the same commutation relation with $\hat{p}$. The quadratic terms in (\ref{expan2.25}) can be seen as a simple harmonic oscillator with the mass $m=\frac{2}{3}$ and  frequency $\omega = \sqrt{3}$, which has the energy spectrum of $\frac{\sqrt{3}(2n+1)\hbar}{2}$ at quantum level $n$. So we find $e^{E^{(n)}} =  3+ \frac{\sqrt{3}(2n+1)\hbar}{2} + \mathcal{O}(\hbar^2) $, and the formula for the first correction is
\begin{eqnarray} \label{E12.26}
E^{(n)}_1 = \frac{ \sqrt{3} (2n+1) }{ 6} .
\end{eqnarray}

Comparing the two formulas (\ref{E12.24}, \ref{E12.26}), we see that the derivative of the classical phase volume at $E_0=\log(3)$ is $ \textrm{vol}_0^{\prime}(E_0) =2\sqrt{3} \pi$. Again we can check numerically that this is indeed an identity using the formula for $\textrm{vol}_0 (E)$ in equation (\ref{classical2.22}).

It turns out that we can not calculate the higher derivatives of classical volume at minimum energy $E_0=\log(3)$ directly with the infinite sum (\ref{classical2.22}). The infinite sum (\ref{classical2.22}) does not converge fast enough at  $E_0=\log(3)$, so that the derivative is not guaranteed to commute with the infinite sum. In practice, we find that the first derivative   $\textrm{vol}_0^{\prime}(E) $ can be still computed numerically by first taking the derivative and then perform the infinite sum. However, for the second derivative, the convergence is slow and the numerical calculation encounters a large error. For the third derivative, the infinite sum becomes divergent at $E_0=\log(3)$.

This is of course not a problem. The $n$-th term in the infinite sum (\ref{classical2.22}) behaves like
\begin{eqnarray}
e^{-3nE}  \frac{(3n-1)!}{n!^3} [\psi(3n) -  \psi(n+1) -E] \sim e^{-3(E- E_0 )n} \frac{E_0 - E} {2\sqrt{3}\pi n^2},
\end{eqnarray}
for large $n$. We see the sum converges rapidly for any $\textrm{Re} (E) > E_0=\log(3)$ and defines the classical volume $\textrm{vol}_0 (E) $ in this domain. We can then analytically continue the classical volume
to the entire complex plane. If the analytic continuation has no pole or cut at $E=E_0$, then all higher derivatives are finite at $E=E_0$.

There are some ways to go about  to compute the higher derivatives at $E=E_0$. We can first compute the derivatives at e.g. $E=E_0+1$, where the derivatives commute with the infinite sum and we can use the formula (\ref{classical2.22}) for numerical calculations. Then we can analytically continue to $E=E_0$  by the Taylor expansion
\begin{eqnarray} \label{Taylor2.29}
\textrm{vol}_0^{(k)}(E_0) = \sum_{n=0}^{\infty}  \frac{(-1)^n \textrm{vol}_0^{(k+n)}(E_0+1)} {n!} .
\end{eqnarray}
We can achieve sufficient numerical accuracy in this way. We check that  the classical volume is indeed analytic at $E=E_0$ and the higher derivatives $\textrm{vol}_0^{(k)}(E_0)$ are finite.

We can also calculate the higher derivatives more effectively using the periods (\ref{conifold2.13}) near the conifold point. Here the classical ground state energy $E=E_0$ corresponds to the conifold point $z=e^{-3E}=\frac{1}{27}$. The classical phase volume $\textrm{vol}_0(E)$ vanishes and has no logarithmic cut at $E=E_0$ , which determines it to be proportional to $t_1(z)$. The constant factor can be also determined by the first derivative $\textrm{vol}^{\prime}_0(E) = 2\sqrt{3}\pi$. We find
\begin{eqnarray}
\textrm{vol}_0(E) = 18\sqrt{3}\pi t_1(z) .
\end{eqnarray}
We can now take derivatives $\partial_E = -3z\partial_z = 3(\frac{1}{27}-z_c) \partial_{z_c}$ repeatedly, and only a finite number of terms in the series expansion in $t_1$ are non-zero when we set $z_c=0$.  In this way we compute the higher derivatives
\begin{eqnarray}\label{low order derivative}
\textrm{vol}_0^{(2)}(E_0) = \frac{4\sqrt{3}\pi}{3}, ~~~
\textrm{vol}_0^{(3)}(E_0) = \frac{4\sqrt{3}\pi}{9}, ~~~~ \textrm{vol}_0^{(4)}(E_0) = - \frac{28\sqrt{3}\pi}{81},
\end{eqnarray}
which have been checked by numerical calculations using (\ref{Taylor2.29}).

The higher order quantum corrections to the phase volume $\textrm{vol}_k(E)$ are related to the leading order one by a second order differential operator \cite{Huang:2012}. We note the convention for Planck constant in \cite{Huang:2012} differs  by a factor of $i$ from here, while the sign for parameter $z$ is opposite. Taking into account the conventions, we have the formulas for the first few orders
\begin{eqnarray} \label{qvolume2.29}
\textrm{vol}_1(E) &=& -\frac{\partial_{E}^2 \textrm{vol}_0 (E)}{72},  \\
  \textrm{vol}_2(E) &=& \frac{-2z(999z+5) \partial_E  \textrm{vol}_0 (E) +z(2619z+29) \partial_E^2  \textrm{vol}_0 (E)  }{1920 \Delta^2},   \nonumber
\end{eqnarray}
where $z=e^{-3E}$ and the discriminant is $\Delta=1-27z$. For the first correction $\textrm{vol}_1(E)$ we can directly plug in the second derivative of classical volume at $E=E_0$. However, for the higher order corrections, e.g. $ \textrm{vol}_2(E)$, we see that there is an apparent pole at $E=E_0$ in the discriminant $\Delta=1-27z$. We should expand both the numerator and denominator around $E\sim E_0$. We find the the final result is finite using the exact values of the derivatives $\textrm{vol}_0^{(k)}(E_0)$. For the first two corrections we find the results
\begin{eqnarray}
\textrm{vol}_1(E_0) = -\frac{\sqrt{3}\pi}{54},  ~~~~ \textrm{vol}_2 (E_0) = \frac{19\sqrt{3}\pi}{209952} .
\end{eqnarray}

With these results we proceed to the higher order energy corrections, where the Bohr-Sommerfeld equation are
\begin{eqnarray} \label{E22.27}
E_2^{(n)} &=&  -\frac{1}{ \textrm{vol}_0^{\prime}(E_0)} [\textrm{vol}_1(E_0) + \frac{(E^{(n) }_1)^2 }{2}  \textrm{vol}_0^{\prime\prime} (E_0)]  \nonumber \\
&=&  -\frac{6n^2+6n+1}{54}, \\
E_3^{(n)}&=& -\frac{E_1^{(n)}\text{vol}_1'(E_0)+E_1^{(n)}E_2^{(n)}\text{vol}_0^{\prime\prime}(E_0)+\frac{1}{6}(E_1^{(n)})^3\text{vol}_0^{\prime\prime\prime}(E_0)}{\text{vol}_0'(E_0)}
\nonumber \\
&=& \frac{10n^3+15n^2+7n+1}{162\sqrt{3}},  \label{third correction}
\end{eqnarray}
where we have used the exact values of classical and quantum phase volume at $E=E_0$.

We can check the higher order corrections through perturbation theory. We expand the Hamiltonian up to order $\hbar^2$ to calculate the second corrections
\begin{equation}
e^{\hat{H}} = 3 + \hat{x}^2 + \hat{p}^2 +\frac{1}{2}\left(\hat{x}\hat{p}+ \hat{p}\hat{x}\right) + \frac{1}{6}\left[ \hat{x}^3 +\hat{p}^3 - ( \hat{x}+\hat{p})^3 \right] + \frac{1}{24}\left[ \hat{x}^4 +\hat{p}^4 + ( \hat{x}+\hat{p})^4 \right]
+\mathcal{O}(\hbar^{\frac{5}{2}}).  \nonumber
\end{equation}
As before we first redefine $\hat{x}^{\prime}=\hat{x}+\frac{\hat{p}}{2}$ to convert the quadratic terms to a simple harmonic oscillator. The creation and annihilation operators can be defined as
\begin{align}\label{xp}
\hat{x}^{\prime}=\frac{3^\frac{1}{4}\sqrt{\hbar}}{2}\left(\hat{a}^\dag+\hat{a}\right),\qquad \hat{p}=\frac{i \sqrt{\hbar}}{3^\frac{1}{4}}\left(\hat{a}^\dag-\hat{a}\right),
\end{align}
satisfying the well-known commutation relation $[\hat{a},\hat{a}^{\dagger}]=1$. By inserting (\ref{xp}) into (\ref{Hamiltonian2.3}), we can express the Hamiltonian as
\begin{align}
e^{\hat{H}}= 3 + \frac{\sqrt{3}\hbar}{2}\left(2\hat{a}^\dag \hat{a}+1\right) +\frac{i\hbar^{\frac{3}{2}}}{2 \cdot 3^{\frac{3}{4}}}\left(\hat{a}\hat{a}\hat{a}-\hat{a}^\dag \hat{a}^\dag \hat{a}^\dag\right)  +\frac{\hbar^2}{8}\left(2\hat{a}^\dag \hat{a} \hat{a}^\dag \hat{a}+2\hat{a}^\dag \hat{a}+1\right)
+\mathcal{O}(\hbar^{\frac{5}{2}}). \nonumber
\end{align}

We use time-independent perturbation theory well-known in quantum mechanics to compute the corrections. See e.g. the textbook \cite{Griffiths}. Define a new Hamiltonian as
\begin{align}
\mathcal{H}=\mathcal{H}_0+\mathcal{H}',
\end{align}
with
\begin{align}
\mathcal{H}_0&=\frac{\sqrt{3}\hbar}{2}\left(2\hat{a}^\dag \hat{a}+1\right),
\\
\mathcal{H}'&= \frac{i\hbar^{\frac{3}{2}}}{2\cdot 3^{\frac{3}{4}}}\left(\hat{a}\hat{a}\hat{a}-\hat{a}^\dag \hat{a}^\dag \hat{a}^\dag\right) + \frac{\hbar^2}{8}\left(2\hat{a}^\dag \hat{a} \hat{a}^\dag \hat{a}+2\hat{a}^\dag \hat{a}+1\right),
\end{align}
where $\mathcal{H}_0$ is the Hamiltonian of a sample harmonic oscillator with the mass $m=\frac{2}{3}$ and frequency $\omega=\sqrt{3}$, and $\mathcal{H}'$ can be treated as a perturbation. The Schr\"{o}dinger equation is
\begin{eqnarray}
\mathcal{H}_0 \psi^{(n)}_0=\mathcal{E}^{(n)}_0 \psi^{(n)}_0,\\
\mathcal{H} \psi^{(n)}=\mathcal{E}^{(n)} \psi^{(n)}, \label{Schrodinger}
\end{eqnarray}
where $\psi^{(n)}_0$ is the wave functions of the harmonic oscillator, and $\mathcal{E}^{(n)}_0=(n+\frac{1}{2})\sqrt{3}\hbar$ is the corresponding energy. It is hard to exactly solve equation (\ref{Schrodinger}). According to the perturbation theory, we can approximately write the solutions as
\begin{align}
\psi^{(n)}&=\psi^{(n)}_0+\psi^{(n)}_1+\cdots =\psi^{(n)}_0+\sum_{m\neq n}\frac{\langle\psi^{(m)}_0|\mathcal{H}'|\psi^{(n)}_0\rangle}{(n-m)\sqrt{3}\hbar}\psi^{(m)}_0+\cdots ,\\
\mathcal{E}^{(n)}&=\mathcal{E}^{(n)}_0+\mathcal{E}^{(n)}_1+\mathcal{E}^{(n)}_2+\cdots
\nonumber
\\
&=(n+\frac{1}{2})\sqrt{3}\hbar+
\langle\psi^{(n)}_0|\mathcal{H}'|\psi^{(n)}_0\rangle+\sum_{m\neq n}\frac{|\langle\psi^{(m)}_0|\mathcal{H}'|\psi^{(n)}_0\rangle|^2}{(n-m)\sqrt{3}\hbar} +\cdots,
\end{align}
where the subscripts denote the different order of the corrections. Using the relations
\begin{align}
\hat{a}|\psi^{(n)}_0\rangle=\sqrt{n}|\psi^{(n-1)}_0\rangle, \qquad \hat{a}^\dag|\psi^{(n)}_0\rangle=\sqrt{n+1}|\psi^{(n+1)}_0\rangle,
\end{align}
it is easy to calculate the energy corrections and give
\begin{align}
\mathcal{E}^{(n)}=(n+\frac{1}{2})\sqrt{3}\hbar+\frac{12n^2+12n+5}{72}\hbar^2+\mathcal{O}(\hbar^3).
\end{align}
So, up to order $\hbar^2$, the eigenvalues of $e^{\hat{H}}$ is $3+(n+\frac{1}{2})\sqrt{3}\hbar+\frac{12n^2+12n+5}{72}\hbar^2$, and eventually gives the second energy spectrum correction
\begin{align}\label{second correction}
E^{(n)}_2=-\frac{6n^2+6n+1}{54},
\end{align}
which does agree with the result (\ref{E22.27}) of the Bohr-Sommerfeld quantization method.

We can also use the time-independent perturbation theory to compute this correction by expanding $e^{\hat{H}}$ to $\hbar^3$ order and calculating $\mathcal{E}^{(n)}_1,\mathcal{E}^{(n)}_2,\mathcal{E}^{(n)}_3,\mathcal{E}^{(n)}_4$. The derivation is too lengthy but similar to lower order calculations, and will not be displayed here. We find the result totally agrees with (\ref{third correction}).

\subsection{Quantum non-perturbative contributions}
\label{subsectionnonper}

In many quantum systems, the perturbative series is a divergent asymptotic series. This is also the case for our model. Of course, the quantum system is well defined for any real value of Planck constant $\hbar$, and one of the key observation of Kallen and Marino in \cite{Kallen:2013} is that the divergence of the perturbative series can be cured by including the non-perturbative contributions. The non-perturbative contributions are usually of the form $e^{-\frac{S_0}{\hbar}}$ where $S_0$ is the action of some instanton configurations. It is difficult to directly calculate the instanton actions. As we mentioned in the introduction, it turns out that in this case the requirement of Kallen-Marino singularity cancellation mechanism largely fix the non-perturbative contributions \cite{Hatsuda:2013, Kallen:2013}.

The perturbative series for our model has singularities when $\hbar$ is a rational number times $\pi$, so the radius of convergence of the perturbative series is actually zero  \cite{Kallen:2013}. When $\hbar$ is small, we can evaluate the quantum spectrum by a truncation of the perturbative series at the minimum term.  Even though the perturbative series is always divergent, the minimum truncation scheme still gives a good approximation to the actual quantum phase volume, with an error of the same order as the minimum term of the series. However, when $\hbar$ is of order one, the non-perturbative contributions become important, and truncating the perturbative series to the first few terms gives not much clue of the actual phase volume.

In order to understand the singularities of the perturbative series, we shall compute the deformed periods exactly in the Planck constant $\hbar$. This is done in the literature \cite{Aganagic:2011}, and we review the calculations here. We denote the deformed A-period and B-period as $\tilde{t}$ and $\tilde{t}_D$, which reduce to the logarithmic and double-logarithmic solutions $w_1(z)$ and $w_2(z)$ in  (\ref{periods2.11}) when $\hbar$ is zero.

We act the curve (\ref{curve2.1}) on a wave function $\psi(x)$ to derive a difference equation
\begin{eqnarray}
(e^x-1)\psi(x) +\psi(x-i\hbar) +ze^{-x-\frac{i\hbar}{2}} \psi(x+ i\hbar) =0.
\end{eqnarray}
Denoting $X=e^x, q=e^{i\hbar}$, and also $V(X) = \frac{\psi(x)}{\psi(x-i\hbar)}$, as in the notation of 
\cite{Hatsuda:2013},  the difference equation is
\begin{eqnarray}
\frac{zV(Xq)}{Xq^{\frac{1}{2}}} +X-1+\frac{1}{V(X)}=0 .
\end{eqnarray}
We can then recursively compute $V(X)$ as a power series of $z$ whose coefficients are exact functions of $\hbar$. The first few terms are
\begin{eqnarray}
V(X) = \frac{1}{1 - X} + \frac{z}{\sqrt{q} X (1 - X)^2  (1 - q X)} + \frac{(1 + q - X - q^3 X) z^2}{
 q^2 X^2 (1 - X)^3  (1 - q X)^2 (1 - q^2 X)}  +\mathcal{O}(z^3). \nonumber
\end{eqnarray}
The power series in the deformed A-period is given by the following residue
\begin{eqnarray}
\tilde{t} &=& \log(z) +3\oint \frac{dx}{2\pi i} \log(V(X))  =\log(z) +3\oint \frac{dX}{2\pi i} \frac{\log(V(X)) }{X} \nonumber \\
&=& \log(z)+ \frac{3 (1 + q) z}{\sqrt{q}} + \frac{3 (2 + 7 q + 12 q^2 + 7 q^3 + 2 q^4) z^2}{ 2 q^2} + (3 + 9 q + 36 q^2 + 88 q^3 + 144 q^4 \nonumber \\  &&
+ 144 q^5 + 88 q^6 +     36 q^7 + 9 q^8 + 3 q^9) \frac{z^3}{q^{9/2}} +\mathcal{O}(z^4) , \label{deformedA2.48}
\end{eqnarray}
where the residue is taken around $X=0$. One can further expand for small $\hbar$ and check the first few order results with formulas (\ref{periods2.11}, \ref{qvolume2.29}).

For the deformed B-period, we need to compute the integral $\int_{\delta}^{\Lambda}  \frac{\log(V(X)) }{X} dX$ with the cut-offs $\delta\sim 0$ and $\Lambda\sim \infty$ in two patches of the local Calabi-Yau geometry \cite{Aganagic:2011}. However, in one of the patches the  recursive process for computing $V(X)$ exactly in $\hbar$ is not so convenient. Instead, we shall use the fact that the  deformed B-period is the derivative of the deformed prepotential, i.e. the Nekrasov-Shatashvili limit of the refined topological string amplitude, with respect to the deformed A-period.

The world-sheet instanton part of the refined topological string amplitude can be written as
\begin{eqnarray} \nonumber
\mathcal{F}_{inst}(t)  \sim \sum_{j_L,j_R} \sum_{m,d=1}^{\infty} \frac{n^{d}_{j_L,j_R}}{m} (-1)^{2j_L+2j_R+md} e^{mdt}
\frac{\sin [m\epsilon_R(2j_R+1)] \sin [m\epsilon_L(2j_L+1)]}{\sin(\frac{m\epsilon_1}{2}) \sin(\frac{m\epsilon_2}{2}) \sin(m\epsilon_R) \sin(m\epsilon_L) }.
\end{eqnarray}
Some explanations of the notations follow. The small parameters $\epsilon_1, \epsilon_2$ parametrize the gravi-photon field strength in 5-dimension by compactifying M-theory on a Calabi-Yau three-fold \cite{IKV}, and the left-right combinations are $\epsilon_{R/L} =\frac{1}{2}(\epsilon_1\pm \epsilon_2)$. The two small parameters are analogous to the ones in $\Omega$-background \cite{Nekrasov}, which is proposed by Nekrasov  to regularize the partition function of Seiberg-Witten theory.  The $n^{d}_{j_L,j_R}$ are the refined version of Gopakumar-Vafa (GV) invariants \cite{GV}, where $j_L, j_R$ are non-negative half integers denoting the spin representations of the 5-dimensional little group $SO(4)\simeq SU(2)_L\times SU(2)_R$.  They are non-negative integers counting numbers of the M2-branes wrapping $d$ times the 2-cycles of Calabi-Yau manifolds. The sum over the integer $m$ denotes the multi-cover contributions.

The refined Gopakumar-Vafa invariants $n^{d}_{j_L,j_R}$ for the local $\mathbb{P}^2$ model are computed by the refined topological vertex \cite{IKV} and also the holomorphic anomaly method in \cite{HK:2010}. A mathematical definition is provided in \cite{Choi:2012}. We list the invariants up to degree $d=7$ in the tables \ref{tableP2} in the Appendix. One salient feature is the ``chess board" pattern. We see that for non-vanishing invariants $n^{d}_{j_L,j_R}$, the sum $2j_L+2j_R+d$ is always an odd integer.

We shall take the Nekrasov-Shatashvili limit, which is
\begin{eqnarray}
\epsilon_1=\hbar,~~~ \epsilon_2\rightarrow 0,~~~ \epsilon_L=\epsilon_R=\frac{\hbar}{2}.
\end{eqnarray}
The world-sheet instanton contributions to the deformed B-period can be computed by the derivative in this limit
\begin{eqnarray} \nonumber
\epsilon_1\epsilon_2 \frac{\partial \mathcal{F}_{inst}(\tilde{t})}{\partial \tilde{t}}
\sim  \sum_{j_L,j_R} \sum_{m,d=1}^{\infty} \frac{2 \hbar d }{m } n^{d}_{j_L,j_R} (-1)^{2j_L+2j_R+md} e^{md\tilde{t}}
\frac{\sin \frac{m\hbar(2j_R+1)}{2} \sin \frac{m\hbar(2j_L+1)}{2}}{\sin^3 (\frac{m\hbar}{2})}.
\end{eqnarray}
The classical contribution to the prepotential is a cubic term $t^3$ from triple intersection of the Calabi-Yau geometry. After fixing the constants, we find the exact $\hbar$ perturbative contribution to the quantum volume of the phase space
\begin{eqnarray} \label{volp2.50}
\textrm{vol}_p(E) &=& \frac{\tilde{t}^2-\pi^2}{2} -\frac{\hbar^2}{8} -\frac{3}{2}   \sum_{j_L,j_R} \sum_{m,d=1}^{\infty} \frac{\hbar d }{m } n^{d}_{j_L,j_R} (-1)^{2j_L+2j_R+md} e^{md\tilde{t}}  \nonumber \\ &&  \times
\frac{\sin \frac{m\hbar(2j_R+1)}{2} \sin \frac{m\hbar(2j_L+1)}{2}}{\sin^3 (\frac{m\hbar}{2})},
\end{eqnarray}
where  the deformed A-period $\tilde{t}$ is available in equation (\ref{deformedA2.48}), and as before $z=e^{-3E}$. Here the constant term $-\frac{\hbar^2}{8}$  is not fixed by the refined GV invariants, it is derived from the first equation in (\ref{qvolume2.29}) when we take the derivatives on the leading double-logarithmic term in the classical phase volume. There is also an extra factor $(-1)^{md}$ comparing to the convention in \cite{IKV, HK:2010}. This is because the convention of $z$ parameter here has opposite sign, as a result the A-period is shifted by a constant of $\pi i$, so the exponent scales as $e^{md t} \rightarrow (-1)^{md} e^{md t}$.  We expand for small $\hbar$ using the refined GV invariants in table \ref{tableP2}, and check the first few order results with formulas (\ref{classical2.22}, \ref{qvolume2.29}).

We can examine the singularities in the perturbative phase volume (\ref{volp2.50}), which comes from the denominator $\sin^3 (\frac{m\hbar}{2})$. It is clear that the singularity appears at $\hbar=\pm \frac{2p\pi }{q}$, where $p,q$ are any two co-prime positive integers. The poles appear when the integer $m$ is an integer  multiple of $q$. We denote $m=m_0 q$, then the pole at $\hbar= \frac{2p\pi }{q}$ is
\begin{eqnarray}
\textrm{vol}_p(E) &=& -3  \sum_{j_L,j_R} \sum_{m_0,d=1}^{\infty} \frac{2\pi p d }{m_0^2 q^3 } n^{d}_{j_L,j_R} (-1)^{2j_L+2j_R+m_0qd} e^{m_0q d\tilde{t}}  \nonumber \\ &&  \times (-1)^{m_0p(2j_L+2j_R+1) }
\frac{(2j_R+1)(2j_L+1) }{\hbar- \frac{2p\pi }{q}} + \mathcal{O}[( \hbar- \frac{2p\pi }{q})^0 ] .
\end{eqnarray}

Certain non-perturbative contributions are proposed in \cite{Hatsuda:2013, Kallen:2013} based on the ordinary, i.e. un-refined, topological string amplitudes, which is the limit
\begin{eqnarray}
\epsilon_1=-\epsilon_2\equiv \epsilon,~~~~ \epsilon_L=\epsilon, ~~~ \epsilon_R\rightarrow 0.
\end{eqnarray}
The topological string amplitude becomes
\begin{eqnarray}
 \mathcal{F}_{inst}(t) \sim  \sum_{j_L,j_R} \sum_{m,d=1}^{\infty} \frac{n^{d}_{j_L,j_R}}{m} (-1)^{2j_L+2j_R+md} e^{mdt}
\frac{(2j_R+1) \sin [m\epsilon (2j_L+1)]}{\sin^2 (\frac{m\epsilon}{2})\sin(m\epsilon ) } .
\end{eqnarray}
In order to cancel the singularities of the perturbative series, we shall take $\epsilon= \frac{4\pi^2}{\hbar}$, and the exponent $e^{mdt }$ is replaced by the non-perturbative form of $e^{\frac{2\pi md t}{\hbar}}$.  We can include some more factors depending only on the product $md$, which do not break the structure of the ordinary topological string amplitude.  After fixing the factors we write the non-perturbative contribution
\begin{eqnarray} \label{volnp}
\textrm{vol}_{np}(E) &=& -\frac{ \hbar }{2} \sum_{j_L,j_R} \sum_{m,d=1}^{\infty} \frac{ n^{d}_{j_L,j_R}}{m} (-1)^{2j_L+2j_R+md} [ \sin (\frac{6\pi^2 md}{\hbar}) e^{\frac{2\pi md \tilde{t} }{\hbar} }  + \cdots ]  \nonumber \\ && \times
\frac{(2j_R+1) \sin [\frac{4\pi^2 m (2j_L+1)}{\hbar}]}{\sin^2 (\frac{2\pi^2 m}{\hbar })\sin(\frac{4\pi^2m}{\hbar} ) } .
\end{eqnarray}
We note that the convention for Planck constant $\hbar$ in \cite{Kallen:2013} is twice of the one here, due to their coordinate transformation. Furthermore, the argument in the $\sin (\frac{6\pi^2 md}{\hbar})$ factor is different. In order to cancel the factor of $(-1)^{md}$ in the perturbative contributions (\ref{volp2.50}), we could have used a factor $\sin (\frac{2k\pi^2 md}{\hbar})$ for any odd integer $k$. It turns out for the local $\mathbb{P}^2$ model, the correct factor is
$\sin (\frac{6\pi^2 md}{\hbar})$. This is not determined by the singularity cancellation requirement, and we shall test its validity with numerical calculations of the spectrum later.

We also write some $\cdots$ in the first line of the above formula (\ref{volnp}) in anticipation of some more smooth corrections. For example, we could add a contribution $\sin^2 (\frac{2 k_1\pi^2 md}{\hbar}) e^{\frac{2k_2\pi md \tilde{t} }{\hbar} } $ in the place of $\cdots$ in the formula, where $k_1, k_2$ are arbitrary integers. This form of correction has no pole for any Planck constant, so it does not affect the singularity cancellation with the perturbative contribution. If the integer $k_2$ is large, then these corrections are quite small, and can only be found by high precision numerical tests.  We will see later that there are indeed such corrections, and we will study them in details in subsection \ref{subsectionhigherorder}.

Similar to the perturbative series, the singularities also appear at $\hbar= \pm \frac{2p\pi }{q}$. Here we denote $m=m_0p$, and the pole at $\hbar=  \frac{2p\pi }{q}$ is
\begin{eqnarray}
\textrm{vol}_{np}(E) &=& 3  \sum_{j_L,j_R} \sum_{m_0,d=1}^{\infty} \frac{2\pi p d }{m_0^2 q^3 } n^{d}_{j_L,j_R} (-1)^{2j_L+2j_R+m_0pd+m_0qd} e^{m_0q d\tilde{t}}  \nonumber \\ &&  \times
\frac{(2j_R+1)(2j_L+1) }{\hbar- \frac{2p\pi }{q}} + \mathcal{O}[( \hbar- \frac{2p\pi }{q})^0 ]  .
 \end{eqnarray}
Since for non-vanishing GV invariants $n^{d}_{j_L,j_R}$ in the local $\mathbb{P}^2$ model, the sum $2j_L+2j_R+d$ is always an odd integer, we find that the poles from perturbative and non-perturbative contributions cancel each others.

The total contribution to the quantum phase volume is then
\begin{eqnarray}
\textrm{vol} (E, \hbar)  = \textrm{vol}_{p}(E) +\textrm{vol}_{np}(E) .
\end{eqnarray}
We consider as examples the some special cases $\hbar=\pi, 2\pi, 3\pi, 5\pi$. Expanding the total quantum phase volume around these points, we find that indeed the poles cancel out. The results of the expansion for large energy up to the first few orders are
\begin{eqnarray} \label{largeE2.55}
\textrm{vol} (E, \pi) &=& \frac{9 E^2}{2} - \frac{5 \pi^2}{8} - \frac{3 \pi}{2} e^{-3E} - \frac{9}{4} (1 + 10 E)  e^{-6E} - \frac{17 \pi}{2}  e^{-9E}  \nonumber \\ && - \frac{9}{16} (7 + 444 E)  e^{-12E} - \frac{1143 \pi}{10} e^{-15E} +\mathcal{O}(e^{-18E}) ,  \\
\label{largeE2.56}
\textrm{vol} (E, 2\pi) &=&  \frac{9 E^2}{2} - \pi^2 + 9 (1 + 5E) e^{-3E} -
 \frac{9}{4} (7 + 222 E) e^{-6E} + (8007 E -188 ) e^{-9E}  \nonumber \\ && +
 \frac{3}{16}(40363 - 797076 E)  e^{-12E} +\mathcal{O}(e^{-15E}),   \\
 \label{largeE2.57}
 \textrm{vol} (E, 3\pi) &=&  \frac{9 E^2}{2} - \frac{13\pi^2}{8}
 + \frac{9 \pi }{2} e^{-3E} - \frac{9(1 + 10 E) }{4} e^{-6E}+ \frac{51 \pi}{2}  e^{-9E}
 \nonumber \\ &&   - \frac{9(7 + 444 E) }{16} e^{-12E} + \frac{ 3429 \pi }{10} e^{-15E} + \mathcal{O}(e^{-18E}), \\
  \label{largeE2.58}
 \textrm{vol} (E, 5\pi) &=&
 \frac{9 E^2}{2} - \frac{29 \pi^2}{8}
 -3\pi \sqrt{5 (5 - 2 \sqrt{5})}  ~e^{-\frac{6}{5}E} +
 \frac{15 \pi }{2} \sqrt{5 - 2 \sqrt{5}} ~e^{-\frac{12}{5}E}  \nonumber \\ &&
 - \frac{15 \pi}{2} e^{-3E} + \mathcal{O}(e^{-\frac{18}{5}E})
\end{eqnarray}

We can solve the energy spectrum in large $E$ expansion. In the leading order we can neglect exponentially small contributions which are powers of $e^{-E}$. We denote the leading order energy $E^{(n)}_0$, which should not be confused with the one in perturbative expansion (\ref{En2.6}) for small $\hbar$.  The Bohr-Sommerfeld condition gives
\begin{eqnarray} \label{leading2.58}
E^{(n)}_0 = \frac{1}{3}[ \pi^2 +\frac{\hbar^2}{4} +2\pi\hbar(2n+1)  ] ^{\frac{1}{2}} .
\end{eqnarray}

The leading order formula (\ref{leading2.58}) is actually a good approximation already. The first exponential correction in the large $E$ expansion is the form $e^{-3 E_0}$ from the perturbative contribution (\ref{volp2.50}) and the form $e^{-\frac{6\pi E_0}{\hbar}}$ from the non-perturbative contribution (\ref{volnp}). So the perturbative contribution dominates over the non-perturbative contribution for $0<\hbar <2\pi$, and vice versa for $h>2\pi$. The first dominant correction is proportional to the greater one of $e^{-3 E_0}$ and $e^{-\frac{6\pi E_0}{\hbar}}$, i.e $\textrm{max}(e^{-3 E_0}, e^{-\frac{6\pi E_0}{\hbar}})$. It is easy to see that the maximum  of $\textrm{max}(e^{-3 E_0}, e^{-\frac{6\pi E_0}{\hbar}})$  is achieved at $\hbar=0$ and $\hbar=\infty$. In both cases, the first exponential correction is proportional to $e^{-\pi} =0.043\ll 1$, so the large $E$ expansion converges well for large or small $\hbar$. On the other hand, for a fixed quantum level $n$, the best convergence occurs at $\hbar=2\pi$, where $\textrm{max}(e^{-3 E_0}, e^{-\frac{6\pi E_0}{\hbar}})$  is at its minimum of $e^{-\pi \sqrt{8n+6}}$.

We use the ansatz for the large $E$ expansion of energy spectrum
\begin{eqnarray}
E^{(n)}(\hbar)  = E^{(n)}_0 +\sum_{j,k=1}^{\infty} c_{j,k} \exp[ - 3(j+\frac{2\pi k }{\hbar} )E^{(n)}_0 ],
\end{eqnarray}
where the exponentials may be the same for different pairs of $(j,k)$ if $\frac{\hbar}{\pi}$ is a rational number, and one should eliminate such redundancies in the sum. We can plug in the large $E$ expansion of the phase volume $\textrm{vol}(E,\hbar)$ and solve for the expansion coefficients $c_{j,k}$ with the Bohr-Sommerfeld quantization condition. We find the results for $\hbar=\pi, 2\pi, 3\pi, 5\pi$ for the fist few terms
\begin{eqnarray}
E^{(n)}(\pi)  &=& E_0 + \frac{\pi e^{-3E_0}}{6E_0} + \frac{180 E^{3}_0+18 E^{ 2}_0  - 6 \pi^2 E_0 - \pi^2 }{72 E^{3}_0}e^{-6E_0}  +\mathcal{O}(e^{-9 E_0}) , \nonumber \\
E^{(n)}(2\pi)  &=& E_0   - \frac{5 E_0+ 1}{E_0}  e^{-3E_0} - \frac{78 E_0^3  + 63 E_0^2 + 12 E_0  +2}{4 E_0^3}  e^{-6E_0}
 +\mathcal{O}(e^{-9E_0}) ,   \nonumber  \\
 E^{(n)}(3\pi)  &=& E_0   - \frac{\pi}{2E_0}  e^{-3E_0} + \frac{20E_0^3+2 E_0^2 -6\pi^2 E_0  -\pi^2 }{8 E_0^3}  e^{-6E_0}
 +\mathcal{O}(e^{-9E_0}) ,   \nonumber  \\
 E^{(n)}(5\pi)  &=& E_0 + \frac{\sqrt{5 (5 - 2 \sqrt{5})} ~\pi }{3 E_0} e^{-\frac{6}{5}E_0} +\mathcal{O}(e^{-\frac{12}{5}E_0}) , \label{largeE2.61}
\end{eqnarray}
where the leading order energy is available in (\ref{leading2.58}), and without confusion of notation we hide the quantum level $n$ by writing $E^{(n)}_0\equiv E_0$. We see that the dependence of the quantum level $n$ only enters through  $E_0$.

\begin{table}
\begin{center}
\begin{tabular} {|c|c|c|} \hline  $E^{(n)}(\hbar=\pi) $    & $n=0$ & $n=1$  \\
\hline  $E_0$ &   \underline{1.887}862233190   &   \underline{2.8196}65699411   \\
\hline  $e^{-3E_0}$ &  \underline{1.8888}24651490   &   \underline{2.819705}063956   \\
 \hline  $e^{-6E_0}$ &  \underline{1.888853}325078   &  \underline{2.8197051753}60  \\
 \hline  $e^{-9E_0}$ &   \underline{1.888853129}661  &   \underline{2.819705175330} \\
 \hline  $e^{-12E_0}$ &   \underline{1.8888531292}75  &  same as above \\
  \hline  $e^{-15E_0}$ &   \underline{1.888853129291}  &  same as above \\  \hline
 \end{tabular}
 \vskip 18pt
\begin{tabular} {|c|c|c|} \hline  $E^{(n)}(\hbar=2\pi) $    & $n=0$ & $n=1$  \\
\hline  $E_0$ &  \underline{2.56}5099660324  &      \underline{3.9182}54452846  \\
\hline  $e^{-3E_0}$ &  \underline{2.56264}7489810   &  \underline{3.91821318}9762  \\
 \hline  $e^{-6E_0}$ &  \underline{2.5626420}82069  &  \underline{3.918213188300}   \\
 \hline  $e^{-9E_0}$ & \underline{2.5626420686}60  &  same as above \\
 \hline  $e^{-12E_0}$ &  \underline{2.562642068624}  & same as above \\
  \hline  $e^{-15E_0}$ &  same as above  &  same as above  \\ \hline
  \end{tabular}
  \vskip 18pt
\begin{tabular} {|c|c|c|} \hline  $E^{(n)}(\hbar=3\pi) $    & $n=0$ & $n=1$  \\
\hline  $E_0$ &  \underline{3.1849}27013119  &       \underline{4.827342}189413  \\
\hline  $e^{-3E_0}$ &  \underline{3.18489206}4364   &   \underline{4.827342022324}  \\
 \hline  $e^{-6E_0}$ &  \underline{3.18489207345}6  &   same as above   \\
 \hline  $e^{-9E_0}$ & \underline{3.184892073458}  &  same as above \\
  \hline
 \end{tabular}
  \vskip 18pt
\begin{tabular} {|c|c|c|} \hline  $E^{(n)}(\hbar=5\pi) $    & $n=0$ & $n=1$  \\
\hline  $E_0$ &  \underline{4.34}9338083980  &      \underline{6.391}337574671  \\
\hline  $e^{-\frac{6}{5}E_0}$ & \underline{4.3514}54881204   &  \underline{6.3914618}30203  \\
 \hline  $e^{-\frac{12}{5}E_0}$ &  \underline{4.35143}6181660  &   \underline{6.39146174}5619   \\
 \hline  $e^{-3E_0}$ & \underline{4.351437}478436  &  \underline{6.39146174}7548 \\
 \hline  $e^{-\frac{18}{5}E_0}$ &  \underline{4.351437}387521  & \underline{6.39146174}7487  \\
  \hline  $e^{-\frac{21}{5}E_0}$ &  \underline{4.351437}375361  & \underline{6.39146174}7486  \\
   \hline  $e^{-\frac{24}{5}E_0}$ &  \underline{4.351437}377729  & same as above  \\
    \hline  $e^{-\frac{27}{5}E_0}$ &  \underline{4.351437}377918  & same as above  \\
     \hline  $e^{-6E_0}$ &  \underline{4.351437}377883  & same as above   \\
      \hline  $e^{-\frac{33}{5}E_0}$ &  same as above   & same as above  \\
  \hline
 \end{tabular}
  \vskip 10pt
\caption{ The energy $E^{(n)}$ from the large $E$ expansion (\ref{largeE2.61}), for the first two quantum levels $n=0,1$, for the cases of $\hbar=\pi, 2\pi, 3\pi, 5\pi$. Each row in the tables denotes the result up to a certain order in the large $E$ expansion. With the knowledge of the  Gopakumar-Vafa invariants up to degree $d$, we can compute the corrections up to (but not include) order $e^{-\textrm{min}(\frac{6\pi^2}{\hbar}, 3)(d+1)E_0}$.  We underline the digits that are checked correctly by the numerical calculations in table \ref{nume}. }
\label{nonper}
 \end{center}
\end{table}

We compute the numerical values of the energy spectrum for the first two quantum levels $n=0,1$, for the cases of $\hbar=\pi, 2\pi, 3\pi, 5\pi$ in the three tables \ref{nonper}.

For the remaining part of this subsection, we consider the limit of large Planck constant $\hbar\rightarrow \infty$. In this case the power series in the perturbative contribution (\ref{volp2.50}) is exponentially small and negligible. According to the formula (\ref{leading2.58}), the energy eigenvalues scale like $E\sim \hbar$, so the higher order terms in the deformed A-period (\ref{deformedA2.48}) are also exponentially small since $z=e^{-3E} \sim e^{-\hbar}$, i.e. we have
 \begin{eqnarray}
 \tilde{t} = \log(z) + \mathcal{O}(e^{-\hbar}) =-3E +\mathcal{O}(e^{-\hbar}) ,  ~~~~ \hbar\rightarrow \infty.
 \end{eqnarray}
If we neglected the non-perturbative contribution, the formula (\ref{leading2.58}) would have been the exact result up to exponentially small corrections in large $\hbar$ limit.  The non-perturbative contribution (\ref{volnp}) scales like $\hbar^2$ and corrects the formula. We can write the first two terms in the large $\hbar$ expansion
\begin{eqnarray} \label{inversehbar2.62}
E(\hbar)  = c_0 \hbar + c_1(\hbar) + \mathcal{O}( \frac{\log(\hbar)^2}{\hbar} ),   ~~~~ \hbar\rightarrow \infty .
\end{eqnarray}
We will see that the leading coefficient $c_0$ is slightly decreased from the naive value of $\frac{1}{6}$ in (\ref{leading2.58}) by the non-perturbative effects. Also it is not a simple power expansion but there will be logarithmic dependence at the sub-leading terms. We have kept the $\hbar$ dependence in the second term $ c_1(\hbar)$ in anticipating of this fact.

Let us determine the first two terms $c_0, c_1(\hbar)$ in the above expansion.  The total quantum phase volume becomes
\begin{eqnarray} \label{total2.64}
\textrm{vol} (E,\hbar) &=&  [\frac{9 E^2}{2\hbar^2}  -\frac{1}{8} -\frac{ 3 }{4\pi^2} \sum_{j_L,j_R} \sum_{m,d=1}^{\infty} \frac{ d }{m^2} n^{d}_{j_L,j_R} (-1)^{2j_L+2j_R+md}   (2j_R+1)  (2j_L+1) \nonumber \\ && \times  e^{ - \frac{6\pi md E }{\hbar}  } ]\hbar^2
 +\mathcal{O} (\hbar^0 ) ,  ~~~~ \hbar\rightarrow \infty.
\end{eqnarray}
Here the sum is exactly the B-period with flat coordinate $-\frac{6\pi E}{\hbar}  $. We shall look for $c_0, c_1(\hbar)$ such that in the above expansion, the coefficient of $\hbar^2$ vanishes and the coefficient of $\hbar$ is $(2n+1)\pi$ according to the Bohr-Sommerfeld quantization condition. We introduce a complex structure parameter $x$ and denote
\begin{eqnarray} \label{relate2.65}
-\frac{6\pi E (\hbar) }{\hbar}= w_1(x),
\end{eqnarray}
where the formula for the A-period $w_1(x)$ is available in (\ref{periods2.11}). In terms of the parameter $x$, the   quantum phase volume (\ref{total2.64}) can be further simply  written as
\begin{eqnarray}
\textrm{vol} (E,\hbar) = [\frac{w_2(x)}{8\pi^2} -\frac{1}{8}] \hbar^2  +\mathcal{O} (\hbar^0 ) ,  ~~~~ \hbar\rightarrow \infty,
\end{eqnarray}
where $w_2(x)$ is the B-period available also in (\ref{periods2.11}). Now we expand around $x\sim \frac{1}{27}$ and use the facts $w_2(\frac{1}{27}) =\pi^2$ and $w_2^{\prime} (\frac{1}{27}) = -36\sqrt{3}\pi$ from the previous subsections, we find
\begin{eqnarray}
\textrm{vol} (E,\hbar) = -\frac{9\sqrt{3}}{2\pi} (x-\frac{1}{27}) \hbar^2   +\mathcal{O} ((x-\frac{1}{27})^2 \hbar^2 ) +\mathcal{O} (\hbar^0 ) ,  ~~~~ \hbar\rightarrow \infty.
\end{eqnarray}
We see that if we identify the parameter
\begin{eqnarray} \label{relate2.68}
\frac{1}{27}-x=\frac{2\pi^2 (2n+1)}{9\sqrt{3}\hbar} + \mathcal{O}(\frac{1}{\hbar^2}) ,
\end{eqnarray}
then the Bohr-Sommerfeld quantization condition is satisfied for the positive $\hbar$ power terms in the quantum phase volume.

When we analytically continue from $x\sim 0$ to $x\sim \frac{1}{27}$,  the A-period $w_1(x)$ is a linear combination that contains the logarithmic solution $t_2$ in (\ref{conifold2.13}). As a result, the $c_1(\hbar)$ is not simply a constant. More precisely, the A-period is actually a hypergeometric function with logarithmic cut at $x\sim \frac{1}{27}$, and the expansion is
\begin{eqnarray} \label{w1expand2.71}
w_1(x) = w_1(\frac{1}{27}) +  \frac{\sqrt{3}(1-27 x)}{2\pi} [\log(\frac{1}{27}-x)-1] + \mathcal{O}[\log(\frac{1}{27}-x) (x- \frac{1}{27})^2],
\end{eqnarray}
where $w_1(\frac{1}{27})=-2.90759 $ is still a finite number. The leading terms determine the full expression as a linear combination of the conifold periods in (\ref{conifold2.13})
\begin{eqnarray}
w_1(x) =  w_1(\frac{1}{27}) + \frac{27\sqrt{3}}{2\pi}[t_2(x)-  t_1(x) ].
\end{eqnarray}
We can plug the relation (\ref{relate2.68}) into the expansion (\ref{w1expand2.71}) and use the relation
(\ref{relate2.65}) to determine $c_0$ and $c_1(\hbar)$ as
\begin{eqnarray} \label{coefficients2.73}
c_0 = - \frac{w_1(\frac{1}{27})}{6\pi}  = 0.154253, ~~~ c_1(\hbar)  = \frac{2n+1}{2} \{ \log[ \frac{9\sqrt{3}\hbar} {2\pi^2(2n+1)}] +1\}.
\end{eqnarray}
We shall test the large $\hbar$ expansion (\ref{inversehbar2.62}) with the above coefficients by numerical calculations in the next subsection.

\subsection{Numerical calculations of the spectrum}
\label{numericalsubsection}

\begin{table}
\begin{center}
\begin{tabular} {|c|c|c|} \hline  $E^{(n)}(\hbar=\pi) $    & $n=0$ & $n=1$  \\
\hline  $100\times 100$ &  \underline{1.888853129}410   &  \underline{2.819705175}780  \\
 \hline  $200\times 200$ & \underline{1.888853129291}   &  \underline{2.819705175330}  \\
  \hline  $300\times 300$ & same as above   &  same as above  \\
 \hline
 \end{tabular}
 \vskip 18pt
\begin{tabular} {|c|c|c|} \hline  $E^{(n)}(\hbar=2\pi) $    & $n=0$ & $n=1$  \\
 \hline  $200\times 200$ & \underline{2.562642068}746   &  \underline{3.918213188}587  \\
 \hline  $300\times 300$ &  \underline{2.562642068624}  & \underline{3.91821318830}1  \\
  \hline  $400\times 400$ &    same as above  & \underline{3.918213188300}   \\
  \hline  $500\times 500$ &    same as above  & same as above   \\
 \hline
 \end{tabular}
  \vskip 18pt
\begin{tabular} {|c|c|c|} \hline  $E^{(n)}(\hbar=3\pi) $    & $n=0$ & $n=1$  \\
 \hline  $200\times 200$ & \underline{3.1848920}89665   &  \underline{4.8273420}52551  \\
 \hline  $300\times 300$ &  \underline{3.184892073}588  & \underline{4.827342022}603  \\
  \hline  $400\times 400$ &    \underline{3.18489207346}1 & \underline{4.82734202232}9   \\
  \hline  $500\times 500$ &     \underline{3.184892073458}  & \underline{4.827342022324}   \\
 \hline
 \end{tabular}
  \vskip 18pt
\begin{tabular} {|c|c|c|} \hline  $E^{(n)}(\hbar=5\pi) $    & $n=0$ & $n=1$  \\
 \hline  $200\times 200$ & \underline{4.3514}48440482   &  \underline{6.3914}78572375  \\
  \hline  $300\times 300$ & \underline{4.351437}967258   &  \underline{6.391462}474309  \\
   \hline  $400\times 400$ & \underline{4.351437}530025   &  \underline{6.3914617}98928  \\
    \hline  $500\times 500$ & \underline{4.351437}500259   &  \underline{6.39146175}2377  \\
 \hline
 \end{tabular}
   \vskip 10pt
\caption{ The energy $E^{(n)}$ from the matrix (\ref{matrix2.64}), for the first two quantum levels $n=0,1$, for the cases of $\hbar=\pi, 2\pi, 3\pi, 5\pi$. Each row in the tables denotes the finite size of the matrix for the eigenvalue computations. We underline the digits that are checked correctly by the large $E$ expansion calculations in table \ref{nonper}. }
\label{nume}
 \end{center}
\end{table}

We shall test the results of the non-perturbative quantum contributions in the previous subsection by direct  numerical  calculations of the quantum spectrum from the Hamiltonian (\ref{Hamiltonian2.3}). A simple choice of the basis is the  wave eigenfunction of the quantum harmonic oscillator with mass $m$ and frequency $w$
\begin{eqnarray} \label{wave2.63}
\psi_n(x) =\frac{1}{\sqrt{2^n n!}} \left( \frac{mw}{\pi \hbar}\right) ^{\frac{1}{4}} e^{-\frac{mwx^2}{2\hbar}} H_n (\sqrt{\frac{mw}{\hbar}} x),
\end{eqnarray}
where $H_n(x)$ are the Hermite polynomials. A useful integral in \cite{GR} is the following
\begin{eqnarray}
\int_{-\infty}^{\infty} e^{-x^2} H_{n_1}(x+y) H_{n_2}(x+z) dx =2^{n_2} \sqrt{\pi} n_1! z^{n_2-n_1} L^{n_2-n_1}_{n_1} (-2yz), ~~~~ n_1 \leq n_2,
\end{eqnarray}
where $L^\alpha_n(z)$ are the Laguerre polynomials.

The action of momentum operator is $e^{\hat{p}}\psi(x) = \psi(x-i\hbar)$. The matrix element can be calculated for $n_1 \leq n_2$ as
\begin{eqnarray} \label{matrix2.64}
\langle \psi_{n_1} | e^{\hat{H}} | \psi_{n_2} \rangle &=& \langle \psi_{n_1} | e^{\hat{x}} + e^{-\frac{\hat{x}}{2} +\hat{p} }
+ e^{-\frac{\hat{x}}{2} -\hat{p} } | \psi_{n_2} \rangle  \nonumber \\
&=&  (\frac{\hbar}{2 mw})^{\frac{n_2-n_1}{2}} \sqrt{\frac{n_1!}{n_2!}} \Big\{ e^{\frac{\hbar}{4mw}} L^{n_2-n_1}_{n_1} (- \frac{\hbar}{2mw}) +  L^{n_2-n_1}_{n_1} (- \frac{\hbar(4m^2w^2+1)}{8mw} ) \nonumber \\
&& \times  e^{\frac{\hbar(4m^2w^2+1)}{16mw}}   [ (-imw -\frac{1}{2}  )^{n_2-n_1} +(imw -\frac{1}{2}  )^{n_2-n_1} ]  \Big\},
\end{eqnarray}
and the matrix element for $n_1 > n_2$  are related by the symmetry $\langle \psi_{n_1} | e^{\hat{H}} | \psi_{n_2} \rangle = \langle \psi_{n_2} | e^{\hat{H}} | \psi_{n_1} \rangle$. Here we have shifted the momentum $\hat{p}\rightarrow \hat{p}-\frac{\hat{x}}{2}$ in the Hamiltonian (\ref{Hamiltonian2.3}) so that the matrix element is real and convenient for numerical calculations. This is somewhat different from the convention in previous subsection \ref{sectionper} where we shifted $\hat{x}$ instead.  We choose the mass and  the frequency $m \omega = \frac{\sqrt{3}}{2}$  from the quadratic term in the small $\hbar$ expansion of the above $e^{\hat{H}}$, which seems to have the best convergence behavior as we increase the matrix size.

We compute the matrix elements $\langle \psi_{n_1} | e^{\hat{H}} | \psi_{n_2} \rangle$ up to some finite level $n$, and compute the eigenvalues of the finite matrix numerically. We expect that when the matrix size is large, the eigenvalues should approach the true quantum energy spectrum asymptotically. The results of the numerical calculations for the first two quantum levels and for the cases of $\hbar=\pi, 2\pi, 3\pi, 5\pi$ are summarized in tables \ref{nume}.  We note that for larger values of $\hbar$, the convergence of the direct numerical calculations from increasing matrix size becomes very slow.

We may also try to improve the convergence by the well known Pad\'{e} approximation. To do this, we compute the energy eigenvalues with increasing sizes with a fixed step. For example, we can use the eigenvalues with matrix sizes $50n\times 50n$, with $n=1,2,\cdots$, up to some finite $n$. The Pad\'{e} approximation can be applied to any finite sequence, and in principle improves its convergence. For more details, see the book \cite{Bender}.

We can compare the results in the tables \ref{nonper} and the tables \ref{nume}. In particular, for the cases $\hbar=\pi, 2\pi, 3\pi$, the two methods converge to the same spectrum and all 12 decimal digits completely agree.  However for the case $\hbar= 5\pi$, the results of the two methods are different starting from the 7th decimal digit. We study the discrepancy in more details in the next subsection, and discover more terms denoted as $\cdots$ in the non-perturbative formula  (\ref{volnp}).

\begin{table}
\begin{center}
\begin{tabular} {|l|l|l|l|l|l| } \hline  $\hbar$ &  $\frac{E^{(0)}(\hbar)}{\hbar} $   & $R^{(3)}_2[R^{(2)}_1[\frac{E^{(0)}(\hbar)}{\hbar} ]]$   & $R^{(3)}_1[R_0[f_1^{(0)}(\hbar) ]]$  &  $R^{(3)}_1[f_2^{(0)}(\hbar) ]$\\
\hline

10 & 0.32952383266224054983 & 0.15375 & 0.49905 & 0.46847\\ \hline11 & 0.3168192283748354037 & 0.15395 & 0.49509 & 0.46783\\ \hline12 & 0.306038429600140043 & 0.15409 & 0.49257 & 0.46700\\ \hline13 & 0.29675930235587053 & 0.15418 & 0.49105 & 0.46610\\ \hline14 & 0.2886768299098925 & 0.15424 & 0.49020 & 0.46520\\ \hline15 & 0.2815647828822417 & 0.15428 & 0.48981 & 0.46433\\ \hline16 & 0.275251591157762 & 0.15430 & 0.48973 & 0.46351\\ \hline17 & 0.269604617679863 & 0.15431 & 0.48986 & 0.46276\\ \hline18 & 0.26451959701770 & 0.15432 & 0.49012 & 0.46206\\ \hline19 & 0.25991335556279 & 0.15432 & 0.49047 & 0.46143\\ \hline20 & 0.2557186778515 & 0.15433 & 0.49087 & 0.46086\\ \hline

 \end{tabular}
  \vskip 18pt

  \begin{tabular}  {|l|l|l|l|l|l| } \hline  $\hbar$ & $\frac{E^{(1)}(\hbar)}{\hbar} $   & $R^{(3)}_2[R^{(2)}_1[\frac{E^{(1)}(\hbar)}{\hbar} ]]$   & $R^{(3)}_1[R_0[f_1^{(1)}(\hbar) ]]$  &  $R^{(3)}_1[f_2^{(1)}(\hbar) ]$\\
\hline

10 & 0.498181382027535340 & 0.15227 & 1.5749 & -0.47827\\ \hline11 & 0.476700618286814275 & 0.15256 & 1.5610 & -0.47247\\ \hline12 & 0.45818227232909232 & 0.15281 & 1.5490 & -0.46821\\ \hline13 & 0.4420182255885214 & 0.15304 & 1.5387 & -0.46512\\ \hline14 & 0.4277607574408575 & 0.15323 & 1.5299 & -0.46293\\ \hline15 & 0.415072099998281 & 0.15340 & 1.5225 & -0.46142\\ \hline16 & 0.403692091900943 & 0.15355 & 1.5163 & -0.46042\\ \hline17 & 0.39341673271950 & 0.15367 & 1.5110 & -0.45982\\ \hline18 & 0.38408355078470 & 0.15377 & 1.5067 & -0.45951\\ \hline19 & 0.3755613681111 & 0.15386 & 1.5030 & -0.45942\\ \hline20 & 0.3677429831319 & 0.15393 & 1.5000 & -0.45950\\ \hline

 \end{tabular}
  \vskip 18pt
  \begin{tabular} {|c|c|c|c|c| } \hline     & $\lim\limits_{\hbar\rightarrow \infty}\frac{E^{(n)}(\hbar)}{\hbar} $   & $ \lim\limits_{\hbar\rightarrow \infty} R_0[f_1^{(n)}(\hbar)]  $  &  $ \lim\limits_{\hbar\rightarrow \infty} f_2^{(n)}(\hbar)  $ \\
  \hline theoretical value & $c_0 =  - \frac{w_1(\frac{1}{27})}{6\pi}$  & $\frac{2n+1}{2}$  &
  $ \frac{2n+1}{2}\{ \log[ \frac{9\sqrt{3}} {2\pi^2(2n+1)}] +1\}$  \\
\hline $n=0$  & 0.154253 & 0.5  & 0.381962 \\
\hline $n=1$  & 0.154253 & 1.5  & -0.502033 \\
 \hline
 \end{tabular}
   \vskip 10pt
\caption{ The Richardson transformations of the  energy spectrum for $n=0,1 $ quantum levels. Here the functions denote $f_1^{(n)}(\hbar)=  E^{(n)}(\hbar) - c_0 \hbar  $ and $f_2^{(n)}(\hbar)  = E^{(n)}(\hbar) - c_0 \hbar -(n+\frac{1}{2})\log(\hbar)$. The theoretical asymptotic values can be found in the formula (\ref{coefficients2.73}). Here for example the transformation $R^{(3)}_2[R^{(2)}_1[\frac{E^{(0)}(\hbar)}{\hbar} ]]$ should eliminate the corrections to $c_0$ up to the form of $\frac{\log(\hbar)^2}{\hbar^2}$. We see that the results of the extrapolation agree well with the theoretical values in $\hbar\rightarrow \infty $ limit for the first two coefficients, while the errors for the last column are somewhat larger.
  }
\label{Richardsontable}
 \end{center}
\end{table}

Now we turn to the numerical test of the large $\hbar$ expansion (\ref{inversehbar2.62}) with the coefficients (\ref{coefficients2.73}). To do this, ideally we should compute the spectrum with very large $\hbar$. However, as mentioned, for a fixed matrix size, the numerical precision in the computation of the spectrum gets worse for larger $\hbar$. It is beyond our computational ability to increase to  the matrix size up to certain level. Instead, we will use the well-known Richardson extrapolation method to test the results (\ref{coefficients2.73}). After some trials, we find that the range $\hbar\sim (10,20)$  provide the best trade-off between larger $\hbar$ and better numerical precision.

Suppose $f(\hbar)$ has the expansion
\begin{eqnarray}
f(\hbar) =f_0 + \frac{f_n}{\hbar^n} +\cdots , ~~~~ \hbar\sim \infty,
\end{eqnarray}
Then we can eliminate the $\hbar^{-n}$ term using the $n$-th order Richardson transformation
\begin{eqnarray}
R_n[f](\hbar)  = \frac{ \hbar^n f(\hbar) - (\hbar - s)^n f(\hbar - s) } {\hbar^n- (\hbar - s)^n },
\end{eqnarray}
where $s$ could be any constant and for simplicity we choose $s=1$.

If there are logarithmic terms in the expansion, e.g.
$f(\hbar) =f_0 + \frac{f_n \log(\hbar) }{\hbar^n} +\cdots$. We can still use the Richardson transformation to eliminate the sub-leading contribution. One can check that doing the transformation twice, i.e. $R^{(2)}_n[f]\equiv R_n[R_n[f]]$, will work. More generally, repeating $(k+1)$-times the $n$-th Richardson transformation $R^{(k+1)}_n[f]$, we can eliminate a sub-leading contribution of the form $\frac{\log(\hbar)^k }{\hbar^n}$.

Furthermore, if $f(\hbar)$ has the logarithmic behavior $f(\hbar) =f_0 \log(\hbar) +f_1 +\cdots$, we can define a 0-th order Richardson transformation
\begin{eqnarray}
R_0[f](\hbar)  = \frac{\hbar[  f(\hbar) -  f(\hbar - s)] } { s } = f_0 +\mathcal{O} (\frac{1}{\hbar}) ,
\end{eqnarray}
which can isolate the coefficient of logarithmic term.

We calculate the energy spectrum $E^{(n)}(\hbar)$ numerically for the integer values of $5\leq \hbar\leq 20$ up to matrix size $900\times 900$ with a step of 50, and also perform a Pad\'{e} approximation to get the energy spectrum closer to the actual values. The results are displayed in tables \ref{Richardsontable}. The expected values in the limit $\hbar\rightarrow \infty$ can be found from the formula (\ref{coefficients2.73}).  We use the Richardson extrapolations  explained above to eliminate some sub-leading corrections, and the results agree well with the expected values.

\subsection{Higher order non-perturbative contributions from precision spectroscopy} \label{subsectionhigherorder}

In this subsection we fix the terms denoted as $\cdots$ in the the non-perturbative formula (\ref{volnp}).  We see  from tables \ref{nonper}, \ref{nume}, the ground state energies for the case of $\hbar=5\pi$ disagree at the 7th decimal digit, which corresponds to the order $e^{- \frac{18}{5}E_0}$ in tables \ref{nonper},  coming from the 3rd sub-leading order of the large $E$ expansion of the non-perturbative contribution. In order to account for the discrepancy, we improve the non-perturbative formula (\ref{volnp})  by the following ansatz
\begin{eqnarray} \label{improvevolnp}
\textrm{vol}_{np}(E) &=& -\frac{ \hbar }{2} \sum_{j_L,j_R} \sum_{m,d=1}^{\infty} \frac{ n^{d}_{j_L,j_R}}{m} (-1)^{2j_L+2j_R+md} [ \sin (\frac{6\pi^2 md}{\hbar}) e^{\frac{2\pi md \tilde{t} }{\hbar} }  + c_3(\frac{ \pi^2 md}{\hbar}) e^{\frac{6\pi md \tilde{t} }{\hbar} }  + \cdots ]  \nonumber \\ && \times
\frac{(2j_R+1) \sin [\frac{4\pi^2 m (2j_L+1)}{\hbar}]}{\sin^2 (\frac{2\pi^2 m}{\hbar })\sin(\frac{4\pi^2m}{\hbar} ) } .
\end{eqnarray}
Here we parametrize the correction $c_3(\frac{ \pi^2 md}{\hbar})$  as  a function of $\frac{md}{\hbar}$, by analogy with the leading term and the exponents. Since the case of $m=d=1$ is the dominant contribution, we can neglect the dependence on the $md$ factor for the first approximation. We should try to determine the exact formula of $c_3(\frac{ \pi^2 }{\hbar})$.

We compute the energy spectrum for the first few quantum levels with Gopakumar-Vafa invariants up to degree $d=7$ and the improved ansatz (\ref{improvevolnp}), using the method in subsection \ref{subsectionnonper}. The error from the actual value is estimated by the last term in the large energy expansion, e.g. in (\ref{largeE2.61}). On the other hand, we also compute the spectrum using the numerical method in subsection \ref{numericalsubsection}. The computation is done with increasing matrix sizes up to $900\times 900$ and we perform a  Pad\'{e} transformation to the sequence. In this case the magnitude of the error from the actual value is estimated by the difference of the last two terms in the converging sequence.

\begin{table}
\begin{center}

 \begin{tabular} {|c|l|c|c|} \hline  & $E^{(n)}(\hbar)$, BS and numerical methods & estimated error  & $c_3(\frac{\pi^2}{\hbar})$  \\ \hline  $\hbar=5, n=0$ & $2.29568495606757869508-3.6500\times 10^{-12}c_3$ & $-3.54\times 10^{-19}$ & $-0.5741$ \\ \hline & $2.29568495606967425812$ & $-1.89\times 10^{-26}$ & \\ \hline $\hbar=5, n=1$ & $3.50180235547641342869-2.8819\times 10^{-18}c_3$ & $-5.99\times 10^{-32}$ & $-0.5740$ \\ \hline & $3.50180235547641343034$ & $-4.45\times 10^{-26}$ & \\ \hline $\hbar=6, n=0$ & $2.50451082107748482697-1.0223\times 10^{-9}c_3$ & $1.41\times 10^{-13}$ & $-0.028016$ \\ \hline & $2.50451082110612706523$ & $-2.10\times 10^{-24}$ & \\ \hline $\hbar=6, n=1$ & $3.82918378903003033292-2.5384\times 10^{-15}c_3$ & $2.21\times 10^{-27}$ & $-0.028287$ \\ \hline & $3.82918378903003040473$ & $-7.41\times 10^{-24}$ & \\ \hline $\hbar=7, n=0$ & $2.70805504957086115105-1.3608\times 10^{-9}c_3$ & $1.36\times 10^{-17}$ & $0.033687$ \\ \hline & $2.70805504952501980724$ & $-1.27\times 10^{-22}$ & \\ \hline $\hbar=7, n=1$ & $4.13741780601652385134-8.6029\times 10^{-15}c_3$ & $3.64\times 10^{-32}$ & $0.032519$ \\ \hline & $4.13741780601652357159$ & $-7.84\times 10^{-22}$ & \\ \hline $\hbar=8, n=0$ & $2.90724838238745761819-1.4014\times 10^{-9}c_3$ & $-1.28\times 10^{-19}$ & $-0.11444$ \\ \hline & $2.90724838254782722204$ & $-3.04\times 10^{-22}$ & \\ \hline $\hbar=8, n=1$ & $4.43050401170032753981-1.9355\times 10^{-14}c_3$ & $4.18\times 10^{-33}$ & $-0.11118$ \\ \hline & $4.43050401170032969180$ & $-1.13\times 10^{-20}$ & \\ \hline $\hbar=9, n=0$ & $3.10279439624530536542-2.5019\times 10^{-9}c_3$ & $-1.10\times 10^{-19}$ & $0.6842$ \\ \hline & $3.10279439453352800511$ & $-1.17\times 10^{-20}$ & \\ \hline $\hbar=9, n=1$ & $4.71127424813225496046-6.7214\times 10^{-14}c_3$ & $-4.99\times 10^{-32}$ & $0.6837$ \\ \hline & $4.71127424813220900419$ & $-1.43\times 10^{-19}$ & \\ \hline $\hbar=10, n=0$ & $3.29523832180536503746-4.8255\times 10^{-9}c_3$ & $-2.31\times 10^{-22}$ & $-0.9982$ \\ \hline & $3.29523832662240549825$ & $-4.61\times 10^{-20}$ & \\ \hline $\hbar=10, n=1$ & $4.98181382027512372478-2.3031\times 10^{-13}c_3$ & $-4.48\times 10^{-33}$ & $-0.9973$ \\ \hline & $4.98181382027535340234$ & $-1.37\times 10^{-18}$ & \\ \hline\end{tabular}

\vskip 10pt
\caption{  We compare the energy spectrum from the two methods, i.e. the Bohr-Sommerfeld and numerical methods.  We compute for many cases of Planck constants, and list some examples in this table. As a consistency check of the calculations, we can see that the solution of $c_3(\frac{ \pi^2 }{\hbar})$ is independent of the quantum level $n$, which only appears on the right hand side of the Bohr-Sommerfeld equation. }
\label{solvec3}
\end{center}
\end{table}

\begin{table}
\begin{center}

 \begin{tabular} {|c|c|c|c|c|c|c|} \hline $\hbar$ & $c_3(\frac{\pi^2}{\hbar}) $ & $\frac{4}{3}\sin^2(\frac{2\pi^2}{\hbar}) \sin(\frac{18\pi^2}{\hbar})$ & ~~& $\hbar$ & $c_3(\frac{\pi^2}{\hbar})$ & $\frac{4}{3}\sin^2(\frac{2\pi^2}{\hbar}) \sin(\frac{18\pi^2}{\hbar})$ \\ \hline$5$ &  $-0.5741$  & $-0.57405$ & &$\pi$ &  $-1.9087\times 10^{-4}$  & $0$ \\ \hline$6$ &  $-0.028016$  & $-0.028291$ & &$2 \pi$ &  $1.0425\times 10^{-10}$  & $0$ \\ \hline$7$ &  $0.033687$  & $0.032493$ & &$3 \pi$ &  $-1.5743\times 10^{-11}$  & $0$ \\ \hline$8$ &  $-0.11444$  & $-0.11109$ & &$4 \pi$ &  $1.3335$  & $1.3333$ \\ \hline$9$ &  $0.6842$  & $0.68371$ & &$\frac{4 \pi }{3}$ &  $-1.3333$  & $-1.3333$ \\ \hline$10$ &  $-0.9982$  & $-0.99723$ & &$5 \pi$ &  $-1.1426$  & $-1.1470$ \\ \hline$11$ &  $-0.5398$  & $-0.54261$ & &$\frac{5 \pi }{2}$ &  $-0.27420$  & $-0.27077$ \\ \hline$12$ &  $1.0509$  & $1.0416$ & &$\frac{5 \pi }{3}$ &  $0.27150$  & $0.27077$ \\ \hline$13$ &  $1.1761$  & $1.1846$ & &$6 \pi$ &  $-1.0141\times 10^{-7}$  & $0$ \\ \hline$14$ &  $0.14676$  & $0.15955$ & &$\frac{7 \pi }{2}$ &  $-0.5472$  & $-0.54987$ \\ \hline$15$ &  $-0.8268$  & $-0.82597$ & &$\frac{7 \pi }{3}$ &  $-0.19387$  & $-0.19625$ \\ \hline$16$ &  $-1.1757$  & $-1.1806$ & &$\frac{8 \pi }{3}$ &  $0.47139$  & $0.47140$ \\ \hline$17$ &  $-0.9564$  & $-0.95905$ & &$\frac{9 \pi }{4}$ &  $1.8126\times 10^{-3}$  & $0$ \\ \hline$18$ &  $-0.45380$  & $-0.45412$ & &$\frac{10 \pi }{3}$ &  $-1.1488$  & $-1.1470$ \\ \hline$19$ &  $0.07384$  & $0.073840$ & &$\frac{15 \pi }{4}$ &  $0.7855$  & $0.77515$ \\ \hline$20$ &  $0.47826$  & $0.47892$ & &$\frac{18 \pi }{5}$ &  $7.060\times 10^{-3}$  & $0$ \\ \hline\end{tabular}

\vskip 10pt
\caption{ We solve the numerical values of $c_3(\frac{ \pi^2 }{\hbar})$ for the ground state level $n=0$ and compare with the conjectured formula for many cases. The agreements provide a convincing test of the formula (\ref{c3formula}). }
\label{checkc3}
\end{center}
\end{table}

The results for some samples of Planck constants and for the first two quantum levels $n=0,1$ are listed in table \ref{solvec3}. The result from the Bohr-Sommerfeld method depends on the function $c_3(\frac{ \pi^2 }{\hbar})$ and as a good approximation we only keep the linear term. If the size of the difference of the energies from the two methods at $c_3=0$ is much bigger than those of the two estimated errors, then there must be significant corrections from the $c_3(\frac{ \pi^2 }{\hbar})$ term to account for the discrepancy, and we can reliably solve for $c_3(\frac{ \pi^2 }{\hbar})$ by equating the results for energy spectrum. Otherwise, the contribution of the $c_3(\frac{ \pi^2 }{\hbar})$ term can not be distinguished from the computational uncertainties. We can still solve for $c_3(\frac{ \pi^2 }{\hbar})$ for some special values of $\hbar$ at which we may suspect $c_3(\frac{ \pi^2 }{\hbar})$ to be zero, and if the solution for $c_3(\frac{ \pi^2 }{\hbar})$ is indeed numerically very close to zero, we may infer that it is actually zero since otherwise its contribution would cause discrepancy unaccounted for by the computational uncertainties.

In order to determine the formula for $c_3(\frac{ \pi^2 }{\hbar})$, we solve for the values of $c_3(\frac{ \pi^2 }{\hbar})$ for many cases of Planck constants $\hbar$ and for the ground state quantum level $n=0$. We choose the values of Planck constant not too small so that the non-perturbative contributions are significant. On the other hand, the Planck constant should not be too large either, so the numerical calculations of matrix eigenvalues do not converge too slowly. We find that the range $5\leq \hbar\leq 20$ is best for the calculations.    After many guesses, we find the correct exact formula
\begin{eqnarray} \label{c3formula}
c_3(\frac{ \pi^2 }{\hbar}) = \frac{4}{3}\sin^2(\frac{2\pi^2}{\hbar}) \sin(\frac{18\pi^2}{\hbar}),
\end{eqnarray}
which agrees well with the numerical solutions of $c_3(\frac{ \pi^2 }{\hbar})$ for all cases of Planck constants. The comparisons are listed in table \ref{checkc3}.

We note that the contribution of the above formula (\ref{c3formula}) to the quantum phase volume (\ref{improvevolnp}) indeed has no singularity for any finite value of Planck constant so it does not spoil the earlier cancellation between non-perturbative and perturbative contributions. Furthermore  this contribution  vanishes for cases of Planck constants when $\frac{18\pi}{\hbar}$ are integers.

Similarly we can proceed to the next orders. Our numerical data are sufficient to help us to guess the following exact formulas for the first few coefficients
\begin{eqnarray} \label{mainresult}
  && \textrm{vol}_{np}(E) =  -\frac{ \hbar }{2} \sum_{j_L,j_R} \sum_{m,d=1}^{\infty} \frac{ n^{d}_{j_L,j_R}}{m} (-1)^{2j_L+2j_R+md} \frac{(2j_R+1) \sin [\frac{4\pi^2 m (2j_L+1)}{\hbar}]}{\sin^2 (\frac{2\pi^2 m}{\hbar })\sin(\frac{4\pi^2m}{\hbar} ) }  \nonumber \\ && ~~~~~~~~ \times  [ \sum_{k=1}^{\infty} c_k(\frac{ \pi^2 md}{\hbar}) e^{\frac{2k \pi md \tilde{t} }{\hbar} }  ], ~~~~~ \textrm{with the following coefficients}
  \nonumber \\ &&   ~~~~ c_1(x) =\sin(6x), ~~~~ ~~~~   c_2(x)=0,
    \nonumber \\ &&   ~~~~c_3(x) = \frac{4}{3}\sin^2(2x) \sin(18x),  ~~~~  c_4(x) = 4 \sin^2(6x) \sin(24x),
 \nonumber \\ &&  ~~~~ c_5(x) = 4\sin(6x)  \sin(30x ) \big[ 7\sin(18x) +16\sin(4x) \sin (6x) \sin (8x)
\nonumber \\ && ~~~~~~~~~~~~ + 4 \sin(2x) \sin(6x) \sin(10x)   \big] ,
\nonumber \\ && ~~~~~~~~~~~~  \cdots .
\end{eqnarray}
Again these next order coefficients consist of at least triple product sine functions, so their contributions are non-singular for any finite value of $\hbar$. Although there is no obvious pattern, it seems that the coefficient  of $e^{\frac{2k \pi md \tilde{t} }{\hbar} }$ always contains a factor of $\sin( \frac{6k \pi^2 md  }{\hbar})$, so it vanishes when $\frac{6k\pi}{\hbar}$ is an integer. If this is true, then in particular, all the higher order contributions vanish when $\frac{6\pi}{\hbar}$ is an integer and the earlier formula (\ref{volnp}) with only the leading term is actually correct in these special cases.

In the large $\hbar$ limit, the higher order contributions to the quantum phase volume go at most like  a constant $\mathcal{O}(\hbar^0)$. So it does not affect the first two coefficients  in the large $\hbar$ expansion of the energy spectrum in equation (\ref{inversehbar2.62}), but will contribute the higher order terms.

\section{The local $\mathbb{P}^1\times \mathbb{P}^1$ model} \label{P1P1section}

This case has been studied in previous literature \cite{Hatsuda:2013, Kallen:2013} for a different formulation relevant for the ABJM matrix model. As we mentioned in the introduction, although the two formulations of Hamiltonian can be related classically by a coordinate transformation, the relation between the quantum theories is more subtle. As such, although we follow the same philosophy, our results for the quantum phase volume and spectrum are different from previous works.

The geometry is described by the classical curve on $(x,p)$ plane
\begin{eqnarray} \label{curve3.1}
e^x + e^p + z_1 e^{-x} +z_2 e^{-p} =1,
\end{eqnarray}
where $z_1,z_2$ are the complex structure modulus parameters of the geometry.

For simplicity one focuses on the $z_1=z_2=z$ case. The Hamiltonian operator is derived from the curve (\ref{curve3.1}) by the following rescaling and shifts
\begin{eqnarray}
z\rightarrow e^{-2H}, ~~~ x\rightarrow x - H, ~~~ p\rightarrow p - H
\end{eqnarray}
Promoting the $x,p$ to the quantum position and momentum operators, we find the one-dimensional quantum mechanical Hamiltonian
\begin{eqnarray} \label{Hamiltonian3.3}
\hat{H} = \log( e^{\hat{x}}+ e^{-\hat{x}}  + e^{\hat{p}} + e^{-\hat{p}} ).
\end{eqnarray}
As in the previous $\mathbb{P}^2$ example, we will compute the perturbative deformed periods using the differential operators in \cite{Huang:2012}. Our method is simpler than that of \cite{Marino:2011, Kallen:2013} for fixing the constant term in the phase volume at order $\hbar^2$.

\subsection{Classical and perturbative contributions}

\begin{figure}[h!]
\begin{center}
\includegraphics[angle=0,width=0.6\textwidth]{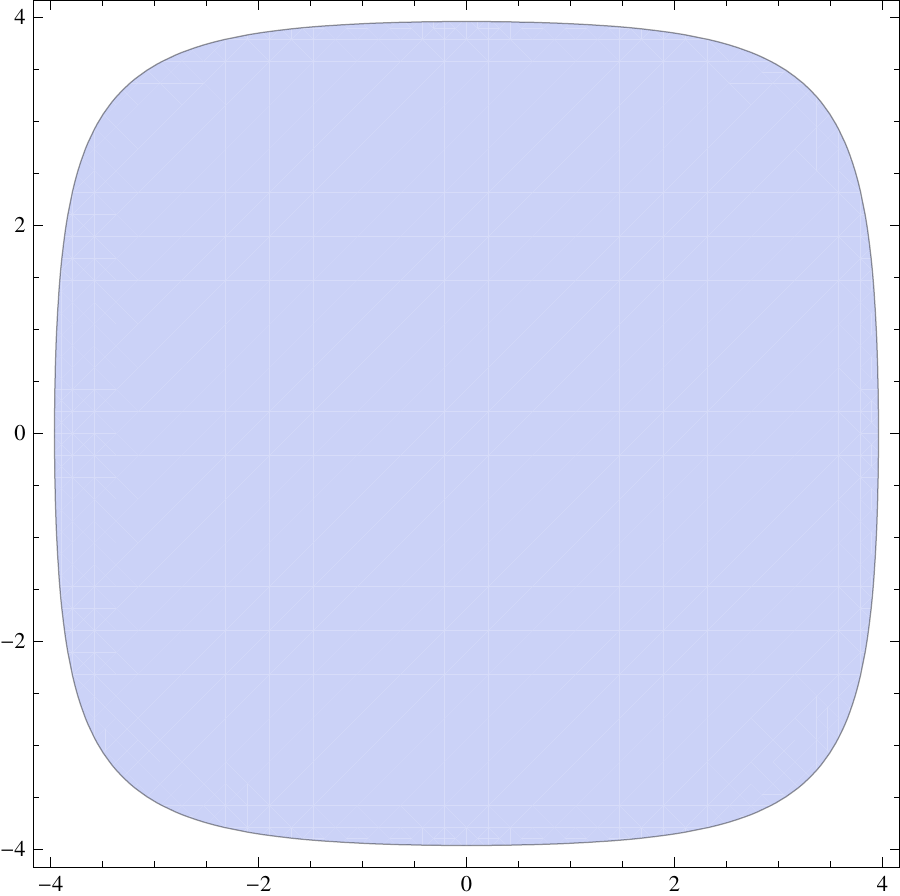}
\begin{quote}
\caption{ The phase space of local $\mathbb{P}^1\times\mathbb{P}^1$ model in the real $(x,p)$ place, parametrized by the equation  $e^x+e^{-x}+e^p +e^{-p} \leq e^E$, for the example of $E=4$.
\vspace{-1.2cm}} \label{figureP1P1}
\end{quote}
\end{center}
\end{figure}

We first compute the perturbative spectrum by the Bohr-Sommerfeld method. The phase space is depicted in Figure \ref{figureP1P1}, which asymptotes to the shape of a square for large $E$. Here for the $z_1=z_2=z$ special case, the relevant Picard-Fuchs differential equation for the $\mathbb{P}^1\times\mathbb{P}^1$ model is
\begin{eqnarray}
[ \Theta_z^3 - 16z(\Theta_z+\frac{1}{2})^2  \Theta_z ] w(z) =0,
\end{eqnarray}
where $\Theta_z =z\partial_z$.  The classical phase volume can be found \cite{Marino:2011} by solving the above equation. The constants are fixed by computing the phase volume in the large energy limit. Here we simply give the result
\begin{eqnarray}
\textrm{vol}_0(E) = 4E^2 -\frac{2\pi^2}{3} + \sum_{n=1}^{\infty} \frac{4}{n} \left(\frac{\Gamma(n+\frac{1}{2})} {\Gamma(\frac{1}{2}) n!}\right)^2 e^{-2n(E-E_0)} [\psi(n+\frac{1}{2}) -\psi(n+1) -\frac{1}{2n}   + E_0-E], \nonumber
\end{eqnarray}
where $E_0=\log(4)$ is the classical ground state energy. One can check numerically $\textrm{vol}_0(E_0)$ vanishes, consistent with the leading order Bohr-Sommerfeld equation.

We compute the derivatives of the classical phase volume at $E=E_0$ and the results are
\begin{eqnarray}
\textrm{vol}_0^{\prime}(E_0) =4\pi, ~~~ \textrm{vol}_0^{\prime\prime}(E_0) =2\pi, ~~~
 \textrm{vol}_0^{(3)}(E_0) =\frac{\pi}{2}, ~~~ \textrm{vol}_0^{(4)}(E_0) =- \frac{\pi}{4}.
\end{eqnarray}

We use the differential operator in \cite{Huang:2012} to compute the quantum correction to the phase volume. The first correction and it derivative are
\begin{eqnarray} \label{op3.8}
\textrm{vol}_1(E_0) &=& - \frac{e^{-2E_0}}{6}  \textrm{vol}_0^{\prime}(E_0) - \frac{1-8 e^{-2E_0}}{48} \textrm{vol}_0^{\prime\prime}(E_0) =- \frac{\pi}{16},  \\ \nonumber
\text{vol}{^\prime}_1(E_0) &=&  \frac{\pi}{64}
\end{eqnarray}
The first few order energy spectrum from the Bohr-Sommerfeld equation is
\begin{eqnarray}
E^{(n)}_1 &=& \frac{ (2n+1)\pi }{ \textrm{vol}_0^{\prime}(E_0)} = \frac{ (2n+1) }{4} , \nonumber \\
E_2^{(n)} &=&  -\frac{1}{ \textrm{vol}_0^{\prime}(E_0)} [\textrm{vol}_1(E_0)  + \frac{(E^{(n) }_1)^2 }{2}  \textrm{vol}_0^{\prime\prime} (E_0)]  =  -\frac{n^2+n}{16},  \nonumber  \\
E_3^{(n)}&= & \frac{10n^3+15n^2+3n-1}{768} .  \label{P1 third correction}
\end{eqnarray}

In \cite{Kallen:2013} the first order quantum phase volume $\textrm{vol}_1(E)$ is written in terms of the complete elliptic integrals. There is a constant contribution in the large $E$ limit, which was calculated in \cite{Marino:2011} using the Wigner approach of quantization. Here we see that the constant is naturally taken into account in the differential operator $(\ref{op3.8})$.

We can again do the calculations in time-independent perturbation theory by expanding $e^{\hat{H}}$ to $\hbar^3$ order and calculating the corresponding corrections $\mathcal{E}^{(n)}_1,\mathcal{E}^{(n)}_2$. Up to order $\hbar^3$,
\begin{align}
e^{\hat{H}}=4 + \hat{x}^2 + \hat{p}^2+\frac{1}{12}\left(\hat{x}^4+\hat{p}^4\right)+\frac{2}{6!}\left(\hat{x}^6+\hat{p}^6\right)+\mathcal{O}(\hbar^{4}).
\end{align}
Similarly to the previous example, we see the quadratic term as a simple harmonic oscillator. The usual creation and annihilation operators are defined as
\begin{eqnarray}
\hat{a} = \frac{\hat{x} +i \hat{p} }{\sqrt{2\hbar}} , ~~~~ \hat{a}^{\dagger} = \frac{\hat{x} - i \hat{p} }{\sqrt{2\hbar}}.
\end{eqnarray}
Treating $\frac{1}{12}\left(\hat{x}^4+\hat{p}^4\right)+\frac{2}{6!}\left(\hat{x}^6+\hat{p}^6\right)$ as perturbation, we can get the corrections to the energy of the harmonic oscillator. We skip the details which are similar to the $\mathbb{P}^2$ model in the previous section. We find the eigenvalues of $e^{\hat{H}}$ and compute the logarithm
\begin{align}
E^{(n)}&=\text{log}\left(4+(2n+1)\hbar+\frac{2n^2+2n+1}{8}\hbar^2+\frac{2n^3+3n^2+3n+1}{192}\hbar^3\right)+\mathcal{O}(\hbar^{4})
\nonumber
\\
&=\text{log}(4)+\frac{2n+1}{4}\hbar-\frac{n^2+n}{16}\hbar^2+\frac{10n^3+15n^2+3n-1}{768}\hbar^3+\mathcal{O}(\hbar^{4}),
\end{align}
which agrees with the energy spectrum (\ref{P1 third correction}) from Bohr-Sommerfeld method.

We should note that the perturbative energy spectrum of the Hamiltonian related classically by a coordinate transformation is also presented in \cite{Kallen:2013} up to second order, quoted as the unpublished work of Hatsuda, Moriyama and Okuyama. Our perturbative method here should be similar, and we present here for the readers' convenience.

\subsection{Non-perturbative contributions}
The exact deformed periods are also calculated in \cite{Aganagic:2011, Hatsuda:2013}. Here we review the calculations for the readers' convenience.  The difference equation for the local $\mathbb{P}^1\times \mathbb{P}^1$ model in the diagonal slice is
\begin{eqnarray}
(e^x+ze^{-x}-1)\psi(x)+\psi(x-i\hbar)+z\psi(x+i\hbar)=0.
\end{eqnarray}
Denoting $X=e^x, q=e^{i\hbar}$, and also $V(X)=\frac{\psi(x)}{\psi(x-i\hbar)}$ as before, the difference equation can be reformulated as
\begin{eqnarray}
(X+\frac{z}{X}-1)+\frac{1}{V(X)}+z V(Xq)=0.
\end{eqnarray}
We still compute $V(X)$ recursively as a power series of $z$ whose coefficients are exact functions of $\hbar$. The result, up to order $z^2$, is
\begin{eqnarray}
V(X) &=& \frac{1}{1-X}+\frac{(q-1)X-1}{(X-1)^2X(qX-1)}z+\frac{z^2 }{ q(X-1)^3X^2(qX-1)^2(q^2X-1)} \nonumber \\ && \times [q-(q^3+2q^2-2q-1)X +(2q^4-q^3-3q^2+2q-1)X^2   \nonumber\\
&& -(q^5-2q^4+q^3-q^2+q)X^3]
+\mathcal{O}(z^3).
\end{eqnarray}
The power series in the deformed A-period is given by the following residue
\begin{eqnarray}
\tilde{t} &=& \text{log}(z) + 2\oint \frac{dx}{2\pi i}\text{log}(V(X))=\text{log}(z)+2\oint \frac{dX}{2\pi i}\frac{\text{log}(V(X))}{X}\label{deformedA3.14}
\\
&=& \text{log}(z)+4z+2(q+\frac{1}{q}+7)z^2+2(2q^2+\frac{2}{q^2}+12q+\frac{12}{q}+\frac{116}{3})z^3+\mathcal{O}(z^4),   \nonumber
\end{eqnarray}
where the residue is taken around $X=0$. One can check this result for small $\hbar$ with the previous formulas.

After fixing the constants, the exact $\hbar$ perturbative contribution to the quantum volume of the phase space
\begin{align}
\text{vol}_p(E)=\tilde{t}^2-\frac{2\pi^2}{3}-\frac{\hbar^2}{6}+\sum_{j_L,j_R}\sum_{m,d=1}^\infty \frac{\hbar d}{m}
n^{d}_{j_L,j_R}e^{m d \tilde{t}}\frac{\sin\frac{m\hbar(2j_L+1)}{2}\sin\frac{m\hbar(2j_R+1)}{2}}{\sin^3\frac{m\hbar}{2}},
\end{align}
where $n^{d}_{j_L,j_R} = \sum_{d_1+d_2=d} n^{d_1,d_2}_{j_L,j_R} $ are the refined Gopakumar-Vafa invariants with $d_1, d_2$ denoting the degrees of the two $\mathbb{P}^1$'s. We sum over the diagonal slice $d=d_1+d_2$ due to the specialization $z_1=z_2$. The invariants have been computed in e.g. \cite{IKV, HK:2010}, and listed here in table \ref{tableP1P1} in the Appendix. Comparing with the formula (\ref{volp2.50}) for local $\mathbb{P}^2$ model, there is no factor of $(-1)^{md}$, since the convention for complex structure parameter $z$ is the same as the one usually used in topological string theory. Furthermore since for the local $\mathbb{P}^1\times \mathbb{P}^1$ model, the non-vanishing GV invariants $n^{d_1,d_2}_{j_L,j_R}$ always have odd integer $2j_L+2j_R$, we can also for simplicity replace the factor $(-1)^{2j_L+2j_R}$ by $-1$.

The poles of the perturbative contributions appear at $\hbar=\frac{2p\pi}{q}$ for integers $p,q$. We denote $m=m_0 q$, then it is
\begin{align}
\text{vol}_p(E)=\sum_{j_L,j_R}\sum_{m_0,d=1}^\infty \frac{4\pi p d}{m_0^2 q^3}n^{d}_{j_L,j_R}e^{m_0 q d \tilde{t}}\frac{(2j_L+1)(2j_R+1)}{\hbar-\frac{2p \pi}{q}}+\mathcal{O}[(\hbar-\frac{2p \pi}{q})^0],
\end{align}
where we have used $(-1)^{m_0 p(2j_L+2j_R+1)}=1$, since $2j_L+2j_R$ is always an odd integer for non-vanishing BPS invariants in the local $\mathbb{P}^1\times \mathbb{P}^1$ model. This is somewhat different from the local $\mathbb{P}^2$ model where $2j_L+2j_R+d$ is always odd instead. 

Similarly we write the non-perturbative contribution as
\begin{align} \label{npP1P1}
\text{vol}_{np}(E)=\sum_{j_L,j_R}\sum_{m,d=1}^\infty \frac{\hbar}{2m}n^{d }_{j_L,j_R}[\sin(\frac{4\pi^2md}{\hbar})
e^{\frac{2\pi m d \tilde{t}}{\hbar}}+\cdots]\frac{(2j_R+1)\sin[\frac{4\pi^2 m(2j_L+1)}{\hbar}]}{\sin^2(\frac{2\pi^2 m}{\hbar})\sin(\frac{4\pi^2 m}{\hbar})}.
\end{align}
We denote $m=m_0 p$, then the pole at $\hbar=\frac{2p\pi}{q}$ is
\begin{align}
\text{vol}_{np}(E)=-\sum_{j_L,j_R}\sum_{m_0,d=1}^\infty \frac{4\pi p d}{m_0^2 q^3}n^{d }_{j_L,j_R}e^{m_0 q d \tilde{t}}\frac{(2j_L+1)(2j_R+1)}{\hbar-\frac{2p \pi}{q}}+\mathcal{O}[(\hbar-\frac{2p \pi}{q})^0],
\end{align}
which exactly cancel the poles from perturbative contribution.

In order to determine the higher order non-perturbative contributions, we calculate the energy spectrum numerically. Again we use the harmonic oscillator basis. The matrix element of the Hamiltonian for $n_1\leqslant n_2$ can be expressed as
\begin{align}
\langle\psi_{n_1}|e^{\hat{H}}|\psi_{n_2}\rangle &= \langle\psi_{n_1}|e^{\hat{x}}+e^{-\hat{x}}+e^{\hat{p}}
+e^{-\hat{p}}|\psi_{n_2}\rangle
\nonumber
\\
&=(\frac{\hbar}{2m\omega})^{\frac{n_2-n_1}{2}}\sqrt{\frac{n_1!}{n_2!}}  [1+(-1)^{n_2-n_1}] \left\{e^{\frac{\hbar}{4m\omega}}L_{n_1}^{n_2-n_1}(-\frac{\hbar}{2m\omega})
\right.
\nonumber
\\
&\left.\quad+(im\omega)^{n_2-n_1} e^{\frac{m\omega\hbar}{4}}L_{n_1}^{n_2-n_1}(-\frac{m\omega\hbar}{2}) \right\},
\end{align}
where we choose the mass $m=\frac{1}{2}$ and the frequency $\omega=2$ as before.

Similarly as the local $\mathbb{P}^2$ model, we compare the energy spectrum from the Bohr-Sommerfeld method and the direct numerical method. We find the first correction to the non-perturbative formula (\ref{npP1P1}) appears at the 4th order. After some high precision calculations, we find the first few order formulas
\begin{eqnarray} \label{mainresultP1P1}
  && \textrm{vol}_{np}(E) =  \frac{ \hbar }{2} \sum_{j_L,j_R} \sum_{m,d=1}^{\infty} \frac{ n^{d}_{j_L,j_R}}{m} \frac{(2j_R+1) \sin [\frac{4\pi^2 m (2j_L+1)}{\hbar}]}{\sin^2 (\frac{2\pi^2 m}{\hbar })\sin(\frac{4\pi^2m}{\hbar} ) }  \nonumber \\ && ~~~~~~~~ \times  [ \sum_{k=1}^{\infty} c_k(\frac{ \pi^2 md}{\hbar}) e^{\frac{2k \pi md \tilde{t} }{\hbar} }  ], ~~~~~ \textrm{with the following coefficients}
  \nonumber \\ &&   ~~~~ c_1(x) =\sin(4x), ~~~~ ~~~~   c_2(x)= c_3(x)=0,
    \nonumber \\ &&   ~~~~c_4(x) = \sin^2(2x) \sin(16x),  ~~~~  c_5(x) = 4 \sin^2(4x) \sin(20x),
  \nonumber \\ &&   ~~~~c_6(x) =  8 \big[ 3 \sin^2(4 x) \sin^2(6 x)+ \sin^2(2 x) \sin^2(8 x)+ \sin^2(10 x) \big] \sin(24 x),
 \nonumber \\ && ~~~~~~~~~~~~  \cdots .
\end{eqnarray}

As in the previous $\mathbb{P}^2$ example, we can provide analytic expansion formulas for the some special cases $\hbar=\pi,2\pi$. This have been done in \cite{Kallen:2013} for the ABJM model related to our convention by a coordinate transformation. The results of the expansion for large energy up to the first few orders are
\begin{align}
\text{vol}(E,\pi)=&4E^2-\frac{5\pi^2}{6}-16Ee^{-2E}+(14-48E)e^{-4E}+(80-\frac{640E}{3})e^{-6E}
\nonumber
\\
&+(\frac{2749}{6}-1128E)e^{-8E}+(2760-\frac{32896E}{5})e^{-10E}+\mathcal{O}(e^{-12E}),
\\
\text{vol}(E,2\pi)=&4E^2-\frac{4\pi^2}{3}-8(1+4E)e^{-2E}-(2+208E)e^{-4E}+\frac{64}{9}(19-276E)e^{-6E}
\nonumber
\\
&+\frac{37}{6}(377-3504E)e^{-8E}+\frac{208}{75}(12197-93360E)e^{-10E}+\mathcal{O}(e^{-12E}).
\end{align}
Note that in these cases, there is no contribution from the higher order corrections in (\ref{mainresultP1P1}) since $\frac{4\pi}{\hbar}$ are integers.

The energy spectrum can also be solved in large $E$ expansion. Neglecting the exponentially small contributions which are powers of $e^{-E}$, we can get the leading order energy $E_0^{(n)}$ by using Bohr-Sommerfeld condition,
\begin{align}\label{leading3.23}
E_0^{(n)}=\frac{1}{2}[\frac{2\pi^2}{3}+\frac{\hbar^2}{6}+(2n+1)\pi\hbar]^{\frac{1}{2}}.
\end{align}
It is easy to find that the first dominant exponential correction is proportional to the greater of $e^{-2E_0},e^{-\frac{4\pi E_0}{\hbar}}$, whose maximum is achieved at $\hbar=0$ and $\hbar=\infty$. In both cases, the first exponential correction is proportional to $e^{-\sqrt{\frac{2}{3}}\pi}=0.077\ll 1$, which ensure that we can reasonably do the large $E$ expansion. Additionally, for a fixed quantum level $n$, the best convergence occurs at $\hbar=2\pi$, where $\textrm{max}(e^{-2 E_0}, e^{-\frac{4\pi E_0}{\hbar}})$  is at its minimum of $e^{-\pi \sqrt{4n+\frac{10}{3}}}$.

We use the ansatz for the large $E$ expansion of energy spectrum
\begin{eqnarray}
E^{(n)}(\hbar)  = E^{(n)}_0 +\sum_{j,k=1}^{\infty} c_{j,k} \exp[ - 2(j+\frac{2\pi k }{\hbar} )E^{(n)}_0 ],
\end{eqnarray}
which is similar to $\mathbb{P}^2$ model.  We give the results for $\hbar=\pi, 2\pi$ for the fist few terms
\begin{eqnarray}
  E^{(n)}(\pi)  &=& E_0+2e^{-2E_0}-\frac{8E_0-1}{4E_0}e^{-4E_0}+\frac{8E_0-3}{3E_0}e^{-6E_0}+\mathcal{O}(e^{-8E_0}),  \\
  E^{(n)}(2\pi) &=& E_0+\frac{4E_0+1}{E_0}e^{-2E_0}-\frac{24E_0^3+31E_0^2+8E_0+2}{4E_0^3}e^{-4E_0}+\mathcal{O}(e^{-6E_0}), 
\end{eqnarray}
where leading order energy is available in (\ref{leading3.23}), and without confusion of notation we hide the quantum level $n$ by writing $E^{(n)}_0\equiv E_0$. We see that the dependence of the quantum level $n$ only enters through  $E_0$.

\begin{table}
\begin{center}
\begin{tabular} {|l|l|l|l|l|l| } \hline  $\hbar$ &  $\frac{E^{(0)}(\hbar)}{\hbar} $   & $R^{(3)}_2[R^{(2)}_1[\frac{E^{(0)}(\hbar)}{\hbar} ]]$   & $R^{(3)}_1[R_0[f_1^{(0)}(\hbar) ]]$  &  $R^{(3)}_1[f_2^{(0)}(\hbar) ]$\\
\hline

10 & 0.370352599041507561767 & 0.18473  & 0.51295  & 0.47393 \\ \hline11 & 0.35632351890932286621 & 0.18508  &  0.50645  & 0.47480 \\ \hline12 & 0.34449075192115692890 &  0.18530  & 0.50215  & 0.47511 \\ \hline13 & 0.3343608759257762637 & 0.18544  & 0.49933  & 0.47511 \\ \hline14 & 0.325579702979615019 & 0.18553  & 0.49751  &0.47492  \\ \hline15 & 0.317886141191162611 & 0.18558   & 0.49636  & 0.47463 \\ \hline16 & 0.3110833112019403 & 0.18561  & 0.49565  & 0.47430 \\ \hline17 & 0.3050198205037168 &  0.18563  & 0.49525  & 0.47395 \\ \hline18 & 0.299577251911793 & 0.18565  & 0.49505  & 0.47360 \\ \hline19 & 0.2946615819217 & 0.18565 &0.49499 & 0.47326 \\ \hline20 & 0.2901971574383 & 0.18565  & 0.49501  & 0.47294 \\ \hline

 \end{tabular}
  \vskip 18pt

  \begin{tabular}  {|l|l|l|l|l|l| } \hline  $\hbar$ & $\frac{E^{(1)}(\hbar)}{\hbar} $   & $R^{(3)}_2[R^{(2)}_1[\frac{E^{(1)}(\hbar)}{\hbar} ]]$   & $R^{(3)}_1[R_0[f_1^{(1)}(\hbar) ]]$  &  $R^{(3)}_1[f_2^{(1)}(\hbar) ]$\\
\hline

10 & 0.5420213334291090387802 & 0.18503  & 1.5550  & -0.46145 \\ \hline11 & 0.51912793976135814591 & 0.18499  & 1.5516  & -0.45710 \\ \hline12 & 0.49945248956002739125 &  0.18501  &  1.5471 & -0.45339 \\ \hline13 & 0.4823252431298052077 & 0.18505  & 1.5427  &-0.45025  \\ \hline14 & 0.467255030845679054 & 0.18510  & 1.5387  & -0.44759 \\ \hline15 & 0.4538725854033678 & 0.18515   & 1.5349  & -0.44534 \\ \hline16 & 0.4418943644298189 & 0.18520  & 1.5315  & -0.44344 \\ \hline17 & 0.431098657645001 & 0.18524  & 1.5285  & -0.44183 \\ \hline18 & 0.421309341378977 & 0.18528  & 1.5257  & -0.44046 \\ \hline19 & 0.412384550474093 & 0.18531 & 1.5232  & -0.43930 \\ \hline20 & 0.404208602584861 & 0.18535  &  1.5209 &-0.43830  \\ \hline

 \end{tabular}
  \vskip 18pt
  \begin{tabular} {|c|c|c|c|c| } \hline     & $\lim\limits_{\hbar\rightarrow \infty}\frac{E^{(n)}(\hbar)}{\hbar} $   & $ \lim\limits_{\hbar\rightarrow \infty} R_0[f_1^{(n)}(\hbar)]  $  &  $ \lim\limits_{\hbar\rightarrow \infty} f_2^{(n)}(\hbar)  $ \\
  \hline theoretical value & $c_0 =  - \frac{w_1(\frac{1}{16})}{4\pi}$  & $\frac{2n+1}{2}$  &
  $ \frac{2n+1}{2}\{ \log[ \frac{8} {\pi^2(2n+1)}] +1\}$  \\
\hline $n=0$  & 0.185614 & 0.5  & 0.394991 \\
\hline $n=1$  & 0.185614 & 1.5  & -0.462946 \\
 \hline
 \end{tabular}
   \vskip 10pt
\caption{ The Richardson transformations of the  energy spectrum for $n=0,1 $ quantum levels for the local $\mathbb{P}^1\times \mathbb{P}^1$ model. Here the functions denote $f_1^{(n)}(\hbar)=  E^{(n)}(\hbar) - c_0 \hbar  $ and $f_2^{(n)}(\hbar)  = E^{(n)}(\hbar) - c_0 \hbar -(n+\frac{1}{2})\log(\hbar)$. The theoretical asymptotic values can be found in the formula (\ref{coefficients3.35}).  We see that the results of the extrapolation agree well with the theoretical values in $\hbar\rightarrow \infty $ limit for the first two coefficients, while the errors for the last column are somewhat larger.
  }
\label{RichardsontableP1}
 \end{center}
\end{table}

Finally we also consider the energy spectrum in the limit of large Planck constant $\hbar\to\infty$ and use Richardson extrapolations to eliminate some sub-leading corrections to compare with theoretical values. Since the method is also the same as $\mathbb{P}^2$ model, we just give the results without detailed explanation. In the limit of large Planck constant $\hbar\to \infty$, we have
\begin{align}
\tilde{t}=\log(z)+\mathcal{O}(e^{-\hbar})=-2E+\mathcal{O}(e^{-\hbar}),
\end{align}
where the energy can be approximately written as
\begin{align}
E(\hbar)=c_0\hbar+c_1(\hbar)+\mathcal{O}(\frac{\log(\hbar)^2}{\hbar}), 
\end{align}
with $c_0,c_1(\hbar)$ will be determined by Bohr-Sommerfeld quantization condition.

The total quantum phase volume becomes
\begin{eqnarray}\label{total3.29}
  \text{vol}(E,\hbar)  &=& [\frac{4E^2}{\hbar^2}-\frac{1}{6}+\frac{1}{2\pi^2}\sum_{j_L,j_R}\sum_{m,d=1}^\infty \frac{d}{m^2}
  n^{d }_{j_L,j_R}(2j_L+1)(2j_R+1)
  \nonumber
  \\
  &&e^{-\frac{4\pi m d E}{\hbar}}]\hbar^2+\mathcal{O}(\hbar^0), \quad \hbar\to \infty.
\end{eqnarray}
Here the sum is exactly the B-period with flat coordinate $-\frac{4\pi E}{\hbar}$. Similarly, we introduce a complex structure parameter $x$ and denote
\begin{align}\label{relation3.30}
-\frac{4\pi E}{\hbar}=\omega_1(x), 
\end{align}
where the formula for the A-period $w_1(x)$ is available in (\ref{deformedA3.14}) by taking $\hbar=0$. In terms of the parameter $x$, the quantum phase volume (\ref{total3.29}) can be further written as
\begin{align}
\text{vol}(E,\hbar)=[\frac{\omega_2(x)}{4\pi^2}-\frac{1}{6}]\hbar^2+\mathcal{O}(\hbar^0), \quad \hbar\to \infty.
\end{align}
By expanding around $x\sim \frac{1}{16}$ and using the facts $\omega_2(\frac{1}{16})=\frac{2\pi^2}{3}$ and $\omega'_2(\frac{1}{16})=-32\pi$, we find
\begin{align}
\text{vol}(E,\hbar)=-\frac{8}{\pi}(x-\frac{1}{16})\hbar^2+\mathcal{O}((x-\frac{1}{16})^2\hbar^2)+\mathcal{O}(\hbar^0), \quad \hbar\to \infty.
\end{align}
The Bohr-Sommerfeld quantization condition gives
\begin{align}\label{relation3.33}
\frac{1}{16}-x=\frac{(2n+1)\pi^2}{8\hbar}+\mathcal{O}(\frac{1}{\hbar^2}).
\end{align}

The expansion around $x\sim \frac{1}{16}$ of $\omega_1(x)$ is
\begin{align}\label{expan3.34}
\omega_1(x)=\omega_1(\frac{1}{16})+\frac{1-16x}{\pi}[\log(\frac{1}{16}-x)-1]+\mathcal{O}[\log(\frac{1}{16}-x)(x-\frac{1}{16})^2],
\end{align}
where $\omega_1(\frac{1}{16})=-2.33249$ is a finite number. Now, we plug the relation (\ref{relation3.33}) into the expansion (\ref{expan3.34}) and use the relation (\ref{relation3.30}) to determine $c_0$ and $c_1(\hbar)$ as
\begin{align}\label{coefficients3.35}
c_0=-\frac{\omega_1(\frac{1}{16})}{4\pi}, \quad c_1(\hbar)=\frac{(2n+1)}{2}[\log(\frac{8\hbar}{\pi^2(2n+1)})+1].
\end{align}
The Richardson extrapolations is displayed in tables \ref{RichardsontableP1}. Again similarly as in the $\mathbb{P}^2$ model, the results agree well with the expected values.

\section{The local $\mathbb{F}_1$ model} \label{F1section}

The local $\mathbb{F}_1$ geometry is a Hirzebruch surface described by the classical curve
\begin{eqnarray} \label{F1geometry}
e^x+z_1 e^{-x} +e^p + z_2 e^{x-p} =1,
\end{eqnarray}
where $z_1, z_2$ are the complex structure moduli parameters, known as the Batyrev coordinates.

According to the studies in \cite{Huang:2013, Huang:2014}, we can construct certain combinations of the Batyrev coordinates, so that only one of the parameters is dynamical and the other parameters can be treated as mass parameters. The quantum period can be computed by the derivatives of only the dynamical parameter. Furthermore, the complex structure moduli space can be seen as a one-dimensional complex plane of the dynamical modulus parameter, so we can solve the topological string amplitudes effectively as one-parameter models and the holomorphic anomaly procedure is greatly simplified.

For the local $\mathbb{F}_1$ model, the correct combination is parametrized as $z_1 = m z^2, z_2 = \frac{z}{m}$. where $z$ is the dynamical parameter and $m$ is the mass parameter. For simplicity we again choose a trivial mass $m=1$. So that the classical curve is
\begin{eqnarray}  \label{geometry3.12}
e^x+z^2 e^{-x} +e^p + z e^{x-p} =1.
\end{eqnarray}

This choice of $z$ parameter is compatible with the derivation of Hamiltonian by the scaling and shifts
\begin{eqnarray}
z\rightarrow e^{-H}, ~~~ x\rightarrow x - H, ~~~ p\rightarrow p - H.
\end{eqnarray}
The quantum Hamiltonian is then
\begin{eqnarray} \label{Hamiltonian3.14}
\hat{H} = \log( e^{\hat{x}}+ e^{-\hat{x}}  + e^{\hat{p}} + e^{\hat{x} -\hat{p}} ).
\end{eqnarray}

\subsection{Classical and perturbative contributions}

\begin{figure}[h!]
\begin{center}
\includegraphics[angle=0,width=0.6\textwidth]{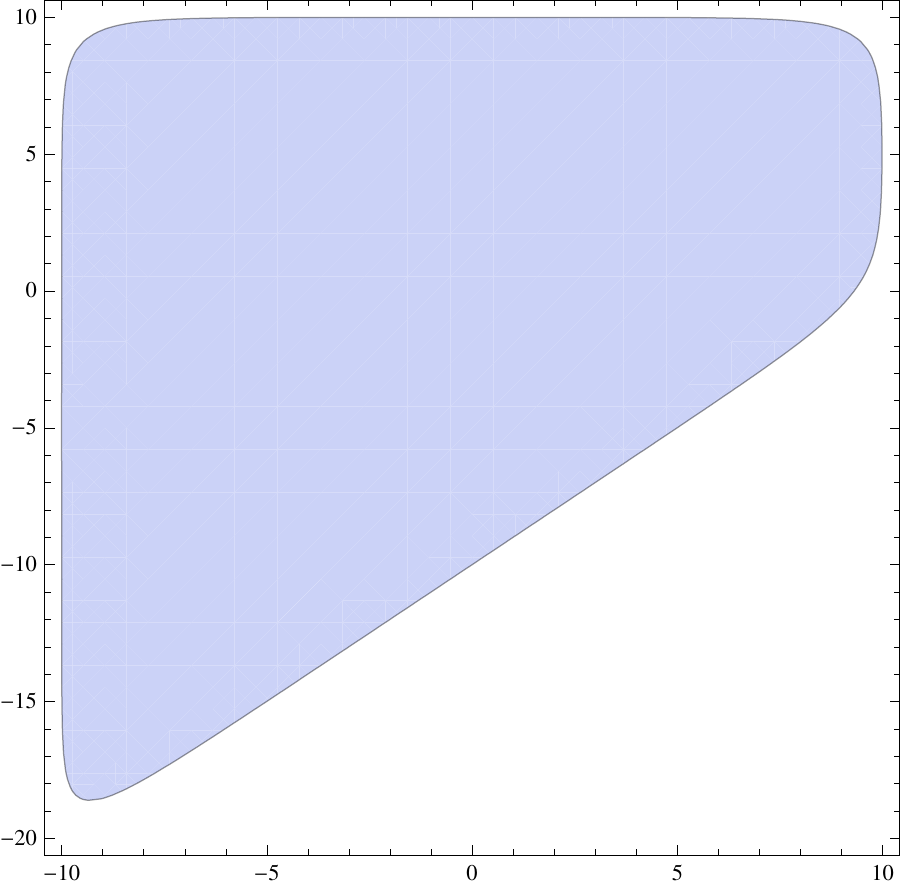}
\begin{quote}
\caption{ The phase space of local $\mathbb{F}_1$ model in the real $(x,p)$ place, parametrized by the equation  $e^x+e^{-x}+e^p +e^{x-p} \leq e^E$, for the example of $E=10$.
\vspace{-1.2cm}} \label{figureF1}
\end{quote}
\end{center}
\end{figure}

The classical phase space is depicted in Figure \ref{figureF1}, which can be seen to asymptote to the shape of a trapezium for large energy.

The classical minimum of the Hamiltonian is achieved at $p=\frac{x}{2}, x=x_0$, where $x_0$ is the only real root of the equation
\begin{eqnarray} \label{x03.15}
e^{x_0}-e^{-x_0} +e^{\frac{x_0}{2}}=0.
\end{eqnarray}
The analytic expression of $x_0$ can be found by solving the above quartic equation for $e^{\frac{x_0}{2}}$, but it is too lengthy to display. Instead we note the numerical value $x_0=-0.3989$.

We can check the perturbative spectrum with Bohr-Sommerfeld method. Here Picard-Fuchs equation is more complicated than the previous example. We can solve for momentum $p$ from the classical geometry (\ref{geometry3.12}), and find the linear combination of the first three derivatives of $z$ that is a total derivative of $x$. In this way we derive the Picard-Fuchs differential equation
\begin{eqnarray} \label{PFeq3.18}
&& [ (8+9z) \Delta(z) \Theta_z^3  -z (1 + 128 z + 936 z^2 + 1000 z^3 + 297 z^4) \Theta_z^2   \\ \nonumber
&&  -2z^2 (32 + 282 z + 294 z^2 + 99 z^3) \Theta_z  ]w(z) =0 ,
\end{eqnarray}
where as before $\Theta_z =z\partial_z$ and the discriminant is
\begin{eqnarray}
 \Delta(z) = 1 + z - 8 z^2 - 36 z^3 - 11 z^4.
\end{eqnarray}
Also as before the discriminant vanishes at the classical minimum, i.e. we have $\Delta(z_0) =0$ for $z_0 = e^{-E_0}$ at the classical minimum $E_0=  \log(2e^{x_0} +3 e^{\frac{x_0}{2}}) = 1.3349 $.

The Picard-Fuchs equation (\ref{PFeq3.18}) is more complicated than the previous cases, and we don't have an analytic expression for the series solutions. Again there are three solutions
\begin{eqnarray}
w_0=1, ~~~  w_1(z) = \log(z) + \sigma_1(z), ~~~ w_2(z)  =\log^2(z)+ 2 \sigma_1 \log(z) +\sigma_2(z),
\end{eqnarray}
where the first few terms of the power series are
\begin{eqnarray}
 \sigma_1(z) &=& z^2 + 2 z^3 + \frac{3}{2} z^4 + 12 z^5 + \frac{55}{3} z^6  + \mathcal{O}(z^7), \nonumber \\
\sigma_2(z) &=&   \frac{z}{4} +  \frac{15}{16} z^2 +   \frac{91}{36} z^3 +  \frac{231}{64} z^4 +  \frac{ 6403}{300} z^5 +  \frac{ 115}{3} z^6 + \mathcal{O}(z^7).   \nonumber
 \end{eqnarray}

We calculate the classical volume $\text{vol}_0(E)$ in large $E$ limit to extract the possible constant contribution from the first period $\omega_0$. The two solutions
\begin{align}
p_\pm(x)=\text{log}\left[\frac{(e^E-e^x-e^{-x})\pm\sqrt{(e^E-e^x-e^{-x})^2-4e^x}}{2}\right]
\end{align}
for the momentum from the Hamiltonian (\ref{Hamiltonian3.14}) at energy $E$ in the classical limit provide a bounded region in the real $(x,p)$ plane and further give the classical volume
\begin{align}\label{F1vol0E}
\text{vol}_0(E)=\int_{e^x+e^{-x}+e^p+e^{x-p}\leqslant e^E}dx dp=\int_a^b \left(p_+(x)-p_-(x)\right)dx,
\end{align}
where the range of the definite integral $a,b$ are the two roots of the equation from the square root term $(e^E-e^x-e^{-x})^2-4e^x=0$, so that $p_+(x)=p_-(x)$ at $x=a,b$, and satisfying $(e^E-e^x-e^{-x})^2-4e^x>0$ for $a<x<b$. This integral is also quite complicated to do exactly, and we imitate the procedure described in the $\mathbb{P}^2$ model. Taking large $E$ limit and neglecting exponentially small corrections, the integration range is then
\begin{align}\label{range}
a=-E+\mathcal{O}(e^{-E}), \qquad  b=E+\mathcal{O}(e^{-E}).
\end{align}
Plugging $p_\pm$ in the phase volume (\ref{F1vol0E}) and substituting the integral range by (\ref{range}), we find
\begin{align}
\text{vol}_0(E) =4E^2+2\int_{-E}^E \text{log}\left[\frac{(1-e^{x-E}-e^{-x-E})+\sqrt{(1-e^{x-E}-e^{-x-E})^2-4e^{x-2E}}}{2}\right] dx. \nonumber
\end{align}
Suppose $ x_0\in (-E, E) $ is a generic value in the integral range, with $x_0+E\sim E-x_0 \sim E $ in the large $E$ limit. We divide the definite integral into two parts, and neglect exponentially small corrections
\begin{align}
\text{vol}_0(E)=4E^2+2\int_{x0}^E \text{log}\left[1-e^{x-E}\right]+2\int_{-E}^{x0} \text{log}\left[1-e^{-x-E}\right] dx.
\end{align}
Using the same techniques as in the $\mathbb{P}^2$ model for the two definite integrals in the above equation, we finally get
\begin{align}
\text{vol}_0(E)=4E^2-\frac{2\pi^2}{3}+\mathcal{O}(e^{-E}).
\end{align}
From the calculations of the phase volume in large $E$, we find the formula for the classical phase volume
\begin{eqnarray}
\text{vol}_0(E) = 4 w_2( e^{-E}) -\frac{2\pi^2 }{3},
\end{eqnarray}
where we replace the variable $z=e^{-E}$ in the B-period. We can check numerically that the classical phase volume vanishes at the minimum $\text{vol}_0(E_0)=0$.

We can compute the derivatives of the classical phase volume at $E_0$ numerically, and the results are the followings
\begin{eqnarray}
\textrm{vol}_0^{\prime}(E_0) = 11.6326 , ~~~ \textrm{vol}_0^{\prime\prime}(E_0) =6.59633 , ~~~
 \textrm{vol}_0^{(3)}(E_0) = 1.67216 .
\end{eqnarray}
The formula for the first few quantum phase volumes has been also obtained in \cite{Huang:2014}. The first correction is
\begin{eqnarray}
\textrm{vol}_1 (E) = -\frac{ 4z^2 (4 + 9 z) \textrm{vol}_0^{\prime}(E)  +   (4 + 3 z - 16 z^2  -36  z^3) \textrm{vol}_0^{\prime\prime}(E)}{24 (8 + 9 z)},
\end{eqnarray}
and the numerical value at classical minimum is $\textrm{vol}_1 (E_0) = -0.162671 $. Basing on this result, we can easily get the first two orders energy spectrum numerically
\begin{eqnarray} \label{BS3.28}
E^{(n)}_1 &=& \frac{ (2n+1)\pi }{ \textrm{vol}_0^{\prime}(E_0)} =  0.54014  (n+\frac{1}{2}) \hbar,   \\
E_2^{(n)} &=&  -\frac{1}{ \textrm{vol}_0^{\prime}(E_0)} [\textrm{vol}_1(E_0) + \frac{(E^{(n) }_1)^2 }{2}  \textrm{vol}_0^{\prime\prime} (E_0)]  \nonumber \\
&=&  -0.0827179(n^2+n) -0.00669548.  \label{BS3.29}
\end{eqnarray}

We can also obtain this spectrum from the perturbation theory, similar to the previous $\mathbb{P}^2$ model.  We first redefine
\begin{align}
\hat{X}=\hat{x}-x_0, \qquad \hat{P}=\hat{p}-\frac{\hat{x}}{2},
\end{align}
which are the small parameters around the classical minimum. The expansion is expressed in $\hat{X},\hat{P}$ below
\begin{align}\label{F1Hamiltonian}
e^{\hat{H}}= &e^{x_0} + e^{-x_0} +2 e^{\frac{x_0}{2}} +\frac{1}{2 m}\hat{P}^2+\frac{1}{2}m \omega^2 \hat{X}^2+\frac{e^{\frac{x_0}{2}}}{3!}\left(\hat{P}^2\hat{X}+\hat{P}\hat{X}\hat{P}+\hat{X}\hat{P}^2-\frac{3}{4}\hat{X}^3\right)
\nonumber
\\
&+\frac{1}{4!}\Big[\frac{e^{\frac{x_0}{2}}}{2}(\hat{P}^2\hat{X}^2+\hat{P}\hat{X}\hat{P}\hat{X}+\hat{P}\hat{X}^2\hat{P}+\hat{X}\hat{P}^2\hat{X}
+\hat{X}\hat{P}\hat{X}\hat{P}+\hat{X}^2\hat{P}^2+4\hat{P}^4-\frac{7}{4}\hat{X}^4)
\nonumber
\\
&+2e^{-x_0}\hat{X}^4\Big]
+\mathcal{O}(\hbar^\frac{5}{2}),
\end{align}
where the mass $m=\frac{1}{2}e^{-\frac{x_0}{2}}$ and the frequency $\omega=\sqrt{ 2e^{\frac{3x_0}{2}} + 2e^{- \frac{x_0}{2}} + e^{x_0} }$. The linear term vanishes since we are expanding around the classical minimum.

We repeat the same procedure as in $\mathbb{P}^2$ model, and get the eigenvalue of $e^{\hat{H}}$ perturbatively up to order $\hbar^2$ as
\begin{align}
e^{E^{(n)}}=&2e^{x_0} + 3e^{\frac{x_0}{2}}+(n+\frac{1}{2})\hbar \omega+\frac{\hbar^2}{256\,m^2\omega^2}\,[e^{\frac{x_0}{2}}(16m^4\omega^4+8m^2\omega^2-7)+16e^{-x_0}]\,(2n^2+2n+1)
\nonumber
\\
&-\frac{e^{x_0}}{512} \frac{\hbar^2 }{m^2 \omega^4}
(-4m^2\omega^2+3)^2(3n^2+3n+1)+(4m^2\omega^2+1)^2(3n^2+3n+2)+\mathcal{O}(\hbar^3), \nonumber
\end{align}
where the $\hbar^2$ term in the first row on the right hand side is the correction from the quartic terms in $e^{\hat{H}}$ and the second row is the correction from cubic terms in $e^{\hat{H}}$. Taking into account the numerical value $x_0=-0.3989$, we finally find the energy spectrum
\begin{align}
E^{(n)}= 1.3349 + 0.54014  (n+\frac{1}{2}) \hbar -[0.0827179(n^2+n)+0.00669548]\hbar^2+\mathcal{O} (\hbar^3),
\end{align}
which obviously agrees with the results (\ref{BS3.28}, \ref{BS3.29}) of Bohr-Sommerfeld method.

\subsection{Non-perturbative contributions}

The difference equation is
\begin{eqnarray}
(e^x+z^2e^{-x}-1)\psi(x)+\psi(x-i\hbar)+ze^{x+\frac{i\hbar}{2}}\psi(x+i\hbar)=0.
\end{eqnarray}
Denoting $X=e^x, q=e^{i\hbar}$, and also $V(X)=\frac{\psi(x)}{\psi(x-i\hbar)}$ as before, the difference equation can be reformulated as
\begin{eqnarray}
(X+\frac{z^2}{X}-1)+\frac{1}{V(X)}+zXq^{\frac{1}{2}}V(Xq)=0.
\end{eqnarray}
We still compute $V(X)$ recursively as a power series of $z$ whose coefficients are exact functions of $\hbar$. The result, up to order $z^2$, is
\begin{eqnarray}
V(X) &=& \frac{1}{1-X}+\frac{\sqrt{q}X}{(X-1)^2(1-qX)}z\nonumber\\
&&+\frac{q(q^2-q-1)X^3-q(q^2+q+1)X^2+(q^2+q+1)X-1}{(X-1)^3X(qX-1)(q^2X-1)}z^2+\mathcal{O}(z^3).
\end{eqnarray}
The power series in the deformed A-period is given by the following residue
\begin{eqnarray}
\tilde{t} &=& \text{log}(z) + \oint \frac{dx}{2\pi i}\text{log}(V(X))=\text{log}(z)+\oint \frac{dX}{2\pi i}\frac{\text{log}(V(X))}{X}
\nonumber
\\
&=& \text{log}(z)+z^2+\frac{(1+q)z^3}{\sqrt{q}}+\frac{3z^4}{2}+\frac{(1+5q+5q^2+q^3)z^5}{q^{\frac{3}{2}}}
\nonumber
\\
&&+\frac{(6+21q+56q^2+21q^3+6q^4)z^6}{6q^2}+O(z^7),
\end{eqnarray}
where the residue is taken around $X=0$. One can check this result for small $\hbar$ with the previous formulas.

The exact $\hbar$ formula for the perturbative contribution to the quantum phase volume is written similarly as previous examples
\begin{eqnarray}
\textrm{vol}_p(E) &=& 4\tilde{t}^2 -\frac{2\pi^2}{3} -\frac{\hbar^2}{6} -\frac{\hbar}{2}   \sum_{j_L,j_R} \sum_{m,d=1}^{\infty} \frac{d }{m } n^{d}_{j_L,j_R} (-1)^{2j_L+2j_R+md} e^{md\tilde{t}}  \nonumber \\ &&  \times
\frac{\sin \frac{m\hbar(2j_R+1)}{2} \sin \frac{m\hbar(2j_L+1)}{2}}{\sin^3 (\frac{m\hbar}{2})},
\end{eqnarray}
where the refined GV invariants $n^{d}_{j_L,j_R} = \sum_{d_B+2d_F=d} n^{d_B,d_F}_{j_L,j_R} $ with the $d_B,d_F$ denoting the degrees of the base $\mathbb{P}^1$ and the fiber $\mathbb{P}^1$.  The combination $d_B+2d_F=d$ is due to our specialization of the complex structure parameters $z_1=z^2, z_2=z$ in the geometry (\ref{F1geometry}). We list the numbers in table \ref{tableF1} in the Appendix. We check the formula with the perturbative calculations in the previous subsection.

In the harmonic oscillator picture, the matrix element of the Hamiltonian for $n_1\leqslant n_2$ can be expressed as
\begin{align}
\langle\psi_{n_1}|e^{\hat{H}}|\psi_{n_2}\rangle &= \langle\psi_{n_1}|e^{x_0}e^{\hat{x}}+e^{-x_0}e^{-\hat{x}}+e^{\frac{x_0}{2}}e^{\frac{\hat{x}}{2}+\hat{p}}
+e^{\frac{x_0}{2}}e^{\frac{\hat{x}}{2}-\hat{p}}|\psi_{n_2}\rangle
\nonumber
\\
&=(\frac{\hbar}{2m\omega})^{\frac{n_2-n_1}{2}}\sqrt{\frac{n_1!}{n_2!}}\left\{e^{\frac{\hbar}{4m\omega}}L_{n_1}^{n_2-n_1}(-\frac{\hbar}{2m\omega})
[e^{x_0}+(-1)^{n_2-n_1}e^{-x_0}]+e^{\frac{x_0}{2}}e^{\frac{\hbar(4m^2\omega^2+1)}{16m\omega}}\right.
\nonumber
\\
&\left.\quad\times L_{n_1}^{n_2-n_1}(-\frac{\hbar(4m^2\omega^2+1)}{8m\omega})
[(-im\omega+\frac{1}{2})^{n_2-n_1}+(im\omega+\frac{1}{2})^{n_2-n_1}]\right\},
\end{align}
where we have performed substitutions $\hat{x}\rightarrow \hat{x}+x_0$ and $\hat{p}\rightarrow \hat{p}+\frac{\hat{x}}{2}+\frac{x_0}{2}$. Note that the mass $m=\frac{1}{2}e^{-\frac{x_0}{2}}$ and the frequency $\omega=\sqrt{2e^{\frac{3x_0}{2}}}+2e^{-\frac{x_0}{2}}+e^{x_0}$ as before.

Similar as the previous examples, we compare the results of the Bohr-Sommerfeld method and direct numerical method. We find the non-perturbative formula with the first few higher order corrections
\begin{eqnarray} \label{mainresultF1}
  && \textrm{vol}_{np}(E) =  - \frac{ \hbar }{2} \sum_{j_L,j_R} \sum_{m,d=1}^{\infty} (-1)^{2j_L+2j_R+md} \frac{ n^{d}_{j_L,j_R}}{m} \frac{(2j_R+1) \sin [\frac{4\pi^2 m (2j_L+1)}{\hbar}]}{\sin^2 (\frac{2\pi^2 m}{\hbar })\sin(\frac{4\pi^2m}{\hbar} ) }  \nonumber \\ && ~~~~~~~~ \times  [ \sum_{k=1}^{\infty} c_k(\frac{ \pi^2 md}{\hbar}) e^{\frac{2k \pi md \tilde{t} }{\hbar} }  ], ~~~~~ \textrm{with the following coefficients}
  \nonumber \\ &&   ~~~~ c_1(x) =\sin(2x), ~~~~ ~~~~   c_2(x)= c_3(x)=\cdots =c_7(x)=0,
  \nonumber \\ &&  ~~~~ c_8(x) = 4\sin^2(2x) \sin(16 x),
 \nonumber \\ && ~~~~~~~~~~~~  \cdots .
\end{eqnarray}
Comparing to the previous examples of the local $\mathbb{P}^2$ and $\mathbb{P}^1\times \mathbb{P}^1$ models, the first non-singular correction appears only at the 8th order and would have been hardly noticeable if we didn't already know its existence.  We see that in all models the first two non-vanishing coefficients have the form $c_1(x)= \sin(2k_1x), c_k(x) \sim \sin^2(2x) \sin(2k_1 kx)$.

\section{Conclusion}

We have considered the spectral problem of a class of quantum Hamiltonians from local Calabi-Yau geometries. We explicitly checked to the first few orders the equivalence of two perturbative methods, namely the time-independent perturbation theory and the Bohr-Sommerfeld method.  In the time-independent perturbation theory, sometimes known as the Rayleigh-Schr\"{o}dinger perturbation theory, we expand the Hamiltonian around the classical minimum. The quadratic term is a simple harmonic oscillator, which can be treated as the zero order term, while the higher order terms are treated as small perturbations. On the other hand, the  Bohr-Sommerfeld quantization condition comes from the consistency condition required by the uniqueness of the quantum mechanical wave function in the well known WKB (Wentzel-Kramers-Brillouin) expansion. Some previous works \cite{Aganagic:2011, Huang:2012, Huang:2014} provide the results of the quantum volume of phase space. It would be interesting to further understand the relation between these two perturbative methods for this class of models.

In the model considered in \cite{Hatsuda:2013, Kallen:2013}, which is essentially the local $\mathbb{P}^1\times \mathbb{P}^1$ model, there is a relation with the ABJM matrix model. It would be interesting to explore whether the other local Calabi-Yau models considered here also have connections with some nice matrix models.

In the well-known example of the quantum mechanical system  with double well potential, the non-perturbative effects come from the instanton sector, which is the solution of the particle going from one minimum to the other one, as reviewed in \cite{Coleman}. Here the non-perturbative contributions to the quantum volume is proposed by the condition that they should cancel the singularities appearing in the perturbative contributions. We also discover more non-singular non-perturbative corrections, in the formulas (\ref{mainresult}, \ref{mainresultP1P1}, \ref{mainresultF1}), by some high precision numerical calculations of the energy spectrum. It would be interesting to understand these non-perturbative contributions directly from instanton configurations of the systems.

\vspace{0.2in} {\leftline {\bf Acknowledgments}}
We thank Albrecht Klemm, Jian-xin Lu and Marcus Marino for discussions and correspondences. MH is supported by the ``Young Thousand People" plan by the Central Organization Department in China, and by the Natural Science Foundation of China.

\appendix
\section{The refined Gopakumar-Vafa invariants}
In this appendix we list the refined Gopakumar-Vafa invariants for the local Calabi-Yau models considered in the paper, in tables \ref{tableP2}, \ref{tableP1P1}, \ref{tableF1}. These invariants are first computed by the refined topological vertex method in \cite{IKV}.

\begin{table}
\begin{center}

  \begin{tabular} {|c|l|} \hline $d$  & $\sum_{j_L,j_R} \oplus n^d_{j_L,j_R}(j_L,j_R)$    \\ \hline 1 & ($0$,$1$)\\ \hline2 & ($0$,$\frac{5}{2}$)\\ \hline3 & ($0$,$3$)$\oplus $($\frac{1}{2}$,$\frac{9}{2}$)\\ \hline4 & ($0$,$\frac{5}{2}$)$\oplus $($0$,$\frac{9}{2}$)$\oplus $($0$,$\frac{13}{2}$)$\oplus $($\frac{1}{2}$,$4$)$\oplus $($\frac{1}{2}$,$5$)$\oplus $($\frac{1}{2}$,$6$)$\oplus $($1$,$\frac{11}{2}$)$\oplus $($\frac{3}{2}$,$7$)\\ \hline5 & ($0$,$1$)$\oplus $($0$,$3$)$\oplus $($0$,$4$)$\oplus $2($0$,$5$)$\oplus $2($0$,$6$)$\oplus $2($0$,$7$)$\oplus $($0$,$8$)$\oplus $($\frac{1}{2}$,$\frac{5}{2}$)$\oplus $($\frac{1}{2}$,$\frac{7}{2}$)$\oplus $2($\frac{1}{2}$,$\frac{9}{2}$)\\ & $\oplus $2($\frac{1}{2}$,$\frac{11}{2}$)$\oplus $3($\frac{1}{2}$,$\frac{13}{2}$)$\oplus $2($\frac{1}{2}$,$\frac{15}{2}$)$\oplus $($\frac{1}{2}$,$\frac{17}{2}$)$\oplus $($1$,$4$)$\oplus $($1$,$5$)$\oplus $2($1$,$6$)$\oplus $2($1$,$7$)$\oplus $2($1$,$8$)\\ & $\oplus $($1$,$9$)$\oplus $($\frac{3}{2}$,$\frac{11}{2}$)$\oplus $($\frac{3}{2}$,$\frac{13}{2}$)$\oplus $2($\frac{3}{2}$,$\frac{15}{2}$)$\oplus $($\frac{3}{2}$,$\frac{17}{2}$)$\oplus $($\frac{3}{2}$,$\frac{19}{2}$)$\oplus $($2$,$7$)$\oplus $($2$,$8$)$\oplus $($2$,$9$)\\ & $\oplus $($\frac{5}{2}$,$\frac{17}{2}$)$\oplus $($3$,$10$)\\ \hline6 & ($0$,$\frac{1}{2}$)$\oplus $($0$,$\frac{3}{2}$)$\oplus $3($0$,$\frac{5}{2}$)$\oplus $2($0$,$\frac{7}{2}$)$\oplus $6($0$,$\frac{9}{2}$)$\oplus $4($0$,$\frac{11}{2}$)$\oplus $8($0$,$\frac{13}{2}$)$\oplus $5($0$,$\frac{15}{2}$)$\oplus $7($0$,$\frac{17}{2}$)\\ & $\oplus $2($0$,$\frac{19}{2}$)$\oplus $2($0$,$\frac{21}{2}$)$\oplus $($\frac{1}{2}$,$1$)$\oplus $2($\frac{1}{2}$,$2$)$\oplus $3($\frac{1}{2}$,$3$)$\oplus $5($\frac{1}{2}$,$4$)$\oplus $6($\frac{1}{2}$,$5$)$\oplus $9($\frac{1}{2}$,$6$)$\oplus $9($\frac{1}{2}$,$7$)\\ & $\oplus $10($\frac{1}{2}$,$8$)$\oplus $7($\frac{1}{2}$,$9$)$\oplus $5($\frac{1}{2}$,$10$)$\oplus $($\frac{1}{2}$,$11$)$\oplus $($\frac{1}{2}$,$12$)$\oplus $($1$,$\frac{3}{2}$)$\oplus $($1$,$\frac{5}{2}$)$\oplus $3($1$,$\frac{7}{2}$)$\oplus $3($1$,$\frac{9}{2}$)\\ & $\oplus $7($1$,$\frac{11}{2}$)$\oplus $7($1$,$\frac{13}{2}$)$\oplus $11($1$,$\frac{15}{2}$)$\oplus $9($1$,$\frac{17}{2}$)$\oplus $9($1$,$\frac{19}{2}$)$\oplus $4($1$,$\frac{21}{2}$)$\oplus $2($1$,$\frac{23}{2}$)$\oplus $($\frac{3}{2}$,$3$)\\ & $\oplus $($\frac{3}{2}$,$4$)$\oplus $3($\frac{3}{2}$,$5$)$\oplus $4($\frac{3}{2}$,$6$)$\oplus $7($\frac{3}{2}$,$7$)$\oplus $7($\frac{3}{2}$,$8$)$\oplus $10($\frac{3}{2}$,$9$)$\oplus $6($\frac{3}{2}$,$10$)$\oplus $4($\frac{3}{2}$,$11$)$\oplus $($2$,$\frac{9}{2}$)\\ & $\oplus $($2$,$\frac{11}{2}$)$\oplus $3($2$,$\frac{13}{2}$)$\oplus $4($2$,$\frac{15}{2}$)$\oplus $7($2$,$\frac{17}{2}$)$\oplus $6($2$,$\frac{19}{2}$)$\oplus $6($2$,$\frac{21}{2}$)$\oplus $2($2$,$\frac{23}{2}$)$\oplus $($2$,$\frac{25}{2}$)\\ & $\oplus $($\frac{5}{2}$,$6$)$\oplus $($\frac{5}{2}$,$7$)$\oplus $3($\frac{5}{2}$,$8$)$\oplus $3($\frac{5}{2}$,$9$)$\oplus $5($\frac{5}{2}$,$10$)$\oplus $3($\frac{5}{2}$,$11$)$\oplus $2($\frac{5}{2}$,$12$)$\oplus $($3$,$\frac{15}{2}$)$\oplus $($3$,$\frac{17}{2}$)\\ & $\oplus $3($3$,$\frac{19}{2}$)$\oplus $3($3$,$\frac{21}{2}$)$\oplus $3($3$,$\frac{23}{2}$)$\oplus $($3$,$\frac{25}{2}$)$\oplus $($\frac{7}{2}$,$9$)$\oplus $($\frac{7}{2}$,$10$)$\oplus $2($\frac{7}{2}$,$11$)$\oplus $($\frac{7}{2}$,$12$)$\oplus $($\frac{7}{2}$,$13$)\\ & $\oplus $($4$,$\frac{21}{2}$)$\oplus $($4$,$\frac{23}{2}$)$\oplus $($4$,$\frac{25}{2}$)$\oplus $($\frac{9}{2}$,$12$)$\oplus $($5$,$\frac{27}{2}$)\\ \hline7 & 6($0$,$1$)$\oplus $6($0$,$2$)$\oplus $12($0$,$3$)$\oplus $13($0$,$4$)$\oplus $19($0$,$5$)$\oplus $21($0$,$6$)$\oplus $26($0$,$7$)$\oplus $26($0$,$8$)\\ & $\oplus $26($0$,$9$)$\oplus $22($0$,$10$)$\oplus $15($0$,$11$)$\oplus $9($0$,$12$)$\oplus $4($0$,$13$)$\oplus $2($0$,$14$)$\oplus $4($\frac{1}{2}$,$\frac{1}{2}$)$\oplus $7($\frac{1}{2}$,$\frac{3}{2}$)\\ & $\oplus $12($\frac{1}{2}$,$\frac{5}{2}$)$\oplus $17($\frac{1}{2}$,$\frac{7}{2}$)$\oplus $24($\frac{1}{2}$,$\frac{9}{2}$)$\oplus $29($\frac{1}{2}$,$\frac{11}{2}$)$\oplus $37($\frac{1}{2}$,$\frac{13}{2}$)$\oplus $41($\frac{1}{2}$,$\frac{15}{2}$)$\oplus $45($\frac{1}{2}$,$\frac{17}{2}$)\\ & $\oplus $41($\frac{1}{2}$,$\frac{19}{2}$)$\oplus $35($\frac{1}{2}$,$\frac{21}{2}$)$\oplus $23($\frac{1}{2}$,$\frac{23}{2}$)$\oplus $13($\frac{1}{2}$,$\frac{25}{2}$)$\oplus $5($\frac{1}{2}$,$\frac{27}{2}$)$\oplus $($\frac{1}{2}$,$\frac{29}{2}$)$\oplus $2($1$,$0$)$\oplus $3($1$,$1$)\\ & $\oplus $8($1$,$2$)$\oplus $11($1$,$3$)$\oplus $18($1$,$4$)$\oplus $23($1$,$5$)$\oplus $33($1$,$6$)$\oplus $40($1$,$7$)$\oplus $48($1$,$8$)$\oplus $50($1$,$9$)\\ & $\oplus $49($1$,$10$)$\oplus $39($1$,$11$)$\oplus $25($1$,$12$)$\oplus $12($1$,$13$)$\oplus $4($1$,$14$)$\oplus $($1$,$15$)$\oplus $($\frac{3}{2}$,$\frac{1}{2}$)$\oplus $3($\frac{3}{2}$,$\frac{3}{2}$)\\ & $\oplus $4($\frac{3}{2}$,$\frac{5}{2}$)$\oplus $9($\frac{3}{2}$,$\frac{7}{2}$)$\oplus $13($\frac{3}{2}$,$\frac{9}{2}$)$\oplus $21($\frac{3}{2}$,$\frac{11}{2}$)$\oplus $27($\frac{3}{2}$,$\frac{13}{2}$)$\oplus $38($\frac{3}{2}$,$\frac{15}{2}$)$\oplus $44($\frac{3}{2}$,$\frac{17}{2}$)$\oplus $50($\frac{3}{2}$,$\frac{19}{2}$)\\ & $\oplus $46($\frac{3}{2}$,$\frac{21}{2}$)$\oplus $38($\frac{3}{2}$,$\frac{23}{2}$)$\oplus $22($\frac{3}{2}$,$\frac{25}{2}$)$\oplus $10($\frac{3}{2}$,$\frac{27}{2}$)$\oplus $3($\frac{3}{2}$,$\frac{29}{2}$)$\oplus $($\frac{3}{2}$,$\frac{31}{2}$)$\oplus $($2$,$1$)$\oplus $($2$,$2$)\\ & $\oplus $3($2$,$3$)$\oplus $5($2$,$4$)$\oplus $10($2$,$5$)$\oplus $14($2$,$6$)$\oplus $22($2$,$7$)$\oplus $29($2$,$8$)$\oplus $38($2$,$9$)$\oplus $41($2$,$10$)\\ & $\oplus $41($2$,$11$)$\oplus $31($2$,$12$)$\oplus $19($2$,$13$)$\oplus $7($2$,$14$)$\oplus $2($2$,$15$)$\oplus $($\frac{5}{2}$,$\frac{5}{2}$)$\oplus $($\frac{5}{2}$,$\frac{7}{2}$)$\oplus $3($\frac{5}{2}$,$\frac{9}{2}$)\\ & $\oplus $5($\frac{5}{2}$,$\frac{11}{2}$)$\oplus $10($\frac{5}{2}$,$\frac{13}{2}$)$\oplus $14($\frac{5}{2}$,$\frac{15}{2}$)$\oplus $22($\frac{5}{2}$,$\frac{17}{2}$)$\oplus $27($\frac{5}{2}$,$\frac{19}{2}$)$\oplus $34($\frac{5}{2}$,$\frac{21}{2}$)$\oplus $32($\frac{5}{2}$,$\frac{23}{2}$)\\ & $\oplus $26($\frac{5}{2}$,$\frac{25}{2}$)$\oplus $14($\frac{5}{2}$,$\frac{27}{2}$)$\oplus $6($\frac{5}{2}$,$\frac{29}{2}$)$\oplus $($\frac{5}{2}$,$\frac{31}{2}$)$\oplus $($3$,$4$)$\oplus $($3$,$5$)$\oplus $3($3$,$6$)$\oplus $5($3$,$7$)$\oplus $10($3$,$8$)\\ & $\oplus $14($3$,$9$)$\oplus $21($3$,$10$)$\oplus $24($3$,$11$)$\oplus $26($3$,$12$)$\oplus $19($3$,$13$)$\oplus $11($3$,$14$)$\oplus $3($3$,$15$)\\ & $\oplus $($3$,$16$)$\oplus $($\frac{7}{2}$,$\frac{11}{2}$)$\oplus $($\frac{7}{2}$,$\frac{13}{2}$)$\oplus $3($\frac{7}{2}$,$\frac{15}{2}$)$\oplus $5($\frac{7}{2}$,$\frac{17}{2}$)$\oplus $10($\frac{7}{2}$,$\frac{19}{2}$)$\oplus $13($\frac{7}{2}$,$\frac{21}{2}$)$\oplus $18($\frac{7}{2}$,$\frac{23}{2}$)\\ & $\oplus $18($\frac{7}{2}$,$\frac{25}{2}$)$\oplus $15($\frac{7}{2}$,$\frac{27}{2}$)$\oplus $7($\frac{7}{2}$,$\frac{29}{2}$)$\oplus $2($\frac{7}{2}$,$\frac{31}{2}$)$\oplus $($4$,$7$)$\oplus $($4$,$8$)$\oplus $3($4$,$9$)$\oplus $5($4$,$10$)\\ & $\oplus $9($4$,$11$)$\oplus $11($4$,$12$)$\oplus $13($4$,$13$)$\oplus $9($4$,$14$)$\oplus $5($4$,$15$)$\oplus $($4$,$16$)$\oplus $($\frac{9}{2}$,$\frac{17}{2}$)$\oplus $($\frac{9}{2}$,$\frac{19}{2}$)\\ & $\oplus $3($\frac{9}{2}$,$\frac{21}{2}$)$\oplus $5($\frac{9}{2}$,$\frac{23}{2}$)$\oplus $8($\frac{9}{2}$,$\frac{25}{2}$)$\oplus $8($\frac{9}{2}$,$\frac{27}{2}$)$\oplus $7($\frac{9}{2}$,$\frac{29}{2}$)$\oplus $3($\frac{9}{2}$,$\frac{31}{2}$)$\oplus $($\frac{9}{2}$,$\frac{33}{2}$)$\oplus $($5$,$10$)\\ & $\oplus $($5$,$11$)$\oplus $3($5$,$12$)$\oplus $4($5$,$13$)$\oplus $6($5$,$14$)$\oplus $4($5$,$15$)$\oplus $2($5$,$16$)$\oplus $($\frac{11}{2}$,$\frac{23}{2}$)$\oplus $($\frac{11}{2}$,$\frac{25}{2}$)\\ & $\oplus $3($\frac{11}{2}$,$\frac{27}{2}$)$\oplus $3($\frac{11}{2}$,$\frac{29}{2}$)$\oplus $3($\frac{11}{2}$,$\frac{31}{2}$)$\oplus $($\frac{11}{2}$,$\frac{33}{2}$)$\oplus $($6$,$13$)$\oplus $($6$,$14$)$\oplus $2($6$,$15$)$\oplus $($6$,$16$)\\ & $\oplus $($6$,$17$)$\oplus $($\frac{13}{2}$,$\frac{29}{2}$)$\oplus $($\frac{13}{2}$,$\frac{31}{2}$)$\oplus $($\frac{13}{2}$,$\frac{33}{2}$)$\oplus $($7$,$16$)$\oplus $($\frac{15}{2}$,$\frac{35}{2}$)\\ \hline\end{tabular}

  \vskip 10pt
\caption{The GV invariants $n^{d}_{j_L,j_R}$ for $d=1,2,\cdots, 7$ for the local $\mathbb{P}^2$ model. }
\label{tableP2}
\end{center}
\end{table}

\begin{table}
\begin{center}

  \begin{tabular} {|c|l|} \hline $d$  & $\sum_{d_1+d_2=d} \sum_{j_L,j_R} \oplus n^{d_1,d_2}_{j_L,j_R}(j_L,j_R)$    \\ \hline 1 & 2($0$,$\frac{1}{2}$)\\ \hline2 & ($0$,$\frac{3}{2}$)\\ \hline3 & 2($0$,$\frac{5}{2}$)\\ \hline4 & ($0$,$\frac{5}{2}$)$\oplus $3($0$,$\frac{7}{2}$)$\oplus $($\frac{1}{2}$,$4$)\\ \hline5 & 2($0$,$\frac{5}{2}$)$\oplus $2($0$,$\frac{7}{2}$)$\oplus $6($0$,$\frac{9}{2}$)$\oplus $2($\frac{1}{2}$,$4$)$\oplus $2($\frac{1}{2}$,$5$)$\oplus $2($1$,$\frac{11}{2}$)\\ \hline6 & ($0$,$\frac{3}{2}$)$\oplus $3($0$,$\frac{5}{2}$)$\oplus $5($0$,$\frac{7}{2}$)$\oplus $7($0$,$\frac{9}{2}$)$\oplus $10($0$,$\frac{11}{2}$)$\oplus $($\frac{1}{2}$,$3$)$\oplus $4($\frac{1}{2}$,$4$)$\oplus $5($\frac{1}{2}$,$5$)$\oplus $7($\frac{1}{2}$,$6$)\\ & $\oplus $($\frac{1}{2}$,$7$)$\oplus $($1$,$\frac{9}{2}$)$\oplus $4($1$,$\frac{11}{2}$)$\oplus $5($1$,$\frac{13}{2}$)$\oplus $($\frac{3}{2}$,$6$)$\oplus $3($\frac{3}{2}$,$7$)$\oplus $($2$,$\frac{15}{2}$)\\ \hline7 & 2($0$,$\frac{1}{2}$)$\oplus $2($0$,$\frac{3}{2}$)$\oplus $8($0$,$\frac{5}{2}$)$\oplus $10($0$,$\frac{7}{2}$)$\oplus $18($0$,$\frac{9}{2}$)$\oplus $16($0$,$\frac{11}{2}$)$\oplus $22($0$,$\frac{13}{2}$)$\oplus $2($0$,$\frac{15}{2}$)\\ & $\oplus $2($0$,$\frac{17}{2}$)$\oplus $2($\frac{1}{2}$,$2$)$\oplus $4($\frac{1}{2}$,$3$)$\oplus $10($\frac{1}{2}$,$4$)$\oplus $14($\frac{1}{2}$,$5$)$\oplus $20($\frac{1}{2}$,$6$)$\oplus $18($\frac{1}{2}$,$7$)$\oplus $4($\frac{1}{2}$,$8$)\\ & $\oplus $2($1$,$\frac{7}{2}$)$\oplus $4($1$,$\frac{9}{2}$)$\oplus $12($1$,$\frac{11}{2}$)$\oplus $14($1$,$\frac{13}{2}$)$\oplus $18($1$,$\frac{15}{2}$)$\oplus $2($1$,$\frac{17}{2}$)$\oplus $2($\frac{3}{2}$,$5$)$\oplus $4($\frac{3}{2}$,$6$)\\ & $\oplus $10($\frac{3}{2}$,$7$)$\oplus $10($\frac{3}{2}$,$8$)$\oplus $2($\frac{3}{2}$,$9$)$\oplus $2($2$,$\frac{13}{2}$)$\oplus $4($2$,$\frac{15}{2}$)$\oplus $8($2$,$\frac{17}{2}$)$\oplus $2($\frac{5}{2}$,$8$)$\oplus $2($\frac{5}{2}$,$9$)\\ & $\oplus $2($3$,$\frac{19}{2}$)\\ \hline8 & 5($0$,$\frac{1}{2}$)$\oplus $11($0$,$\frac{3}{2}$)$\oplus $19($0$,$\frac{5}{2}$)$\oplus $30($0$,$\frac{7}{2}$)$\oplus $40($0$,$\frac{9}{2}$)$\oplus $50($0$,$\frac{11}{2}$)$\oplus $49($0$,$\frac{13}{2}$)$\oplus $50($0$,$\frac{15}{2}$)\\ & $\oplus $14($0$,$\frac{17}{2}$)$\oplus $8($0$,$\frac{19}{2}$)$\oplus $4($\frac{1}{2}$,$1$)$\oplus $9($\frac{1}{2}$,$2$)$\oplus $16($\frac{1}{2}$,$3$)$\oplus $31($\frac{1}{2}$,$4$)$\oplus $44($\frac{1}{2}$,$5$)$\oplus $60($\frac{1}{2}$,$6$)\\ & $\oplus $64($\frac{1}{2}$,$7$)$\oplus $57($\frac{1}{2}$,$8$)$\oplus $20($\frac{1}{2}$,$9$)$\oplus $5($\frac{1}{2}$,$10$)$\oplus $($1$,$\frac{3}{2}$)$\oplus $4($1$,$\frac{5}{2}$)$\oplus $10($1$,$\frac{7}{2}$)$\oplus $20($1$,$\frac{9}{2}$)\\ & $\oplus $36($1$,$\frac{11}{2}$)$\oplus $52($1$,$\frac{13}{2}$)$\oplus $60($1$,$\frac{15}{2}$)$\oplus $55($1$,$\frac{17}{2}$)$\oplus $14($1$,$\frac{19}{2}$)$\oplus $4($1$,$\frac{21}{2}$)$\oplus $($\frac{3}{2}$,$3$)$\oplus $4($\frac{3}{2}$,$4$)\\ & $\oplus $10($\frac{3}{2}$,$5$)$\oplus $20($\frac{3}{2}$,$6$)$\oplus $36($\frac{3}{2}$,$7$)$\oplus $44($\frac{3}{2}$,$8$)$\oplus $44($\frac{3}{2}$,$9$)$\oplus $12($\frac{3}{2}$,$10$)$\oplus $($\frac{3}{2}$,$11$)$\oplus $($2$,$\frac{9}{2}$)\\ & $\oplus $4($2$,$\frac{11}{2}$)$\oplus $10($2$,$\frac{13}{2}$)$\oplus $20($2$,$\frac{15}{2}$)$\oplus $31($2$,$\frac{17}{2}$)$\oplus $31($2$,$\frac{19}{2}$)$\oplus $5($2$,$\frac{21}{2}$)$\oplus $($\frac{5}{2}$,$6$)$\oplus $4($\frac{5}{2}$,$7$)\\ & $\oplus $10($\frac{5}{2}$,$8$)$\oplus $16($\frac{5}{2}$,$9$)$\oplus $19($\frac{5}{2}$,$10$)$\oplus $4($\frac{5}{2}$,$11$)$\oplus $($3$,$\frac{15}{2}$)$\oplus $4($3$,$\frac{17}{2}$)$\oplus $9($3$,$\frac{19}{2}$)$\oplus $11($3$,$\frac{21}{2}$)\\ & $\oplus $($3$,$\frac{23}{2}$)$\oplus $($\frac{7}{2}$,$9$)$\oplus $4($\frac{7}{2}$,$10$)$\oplus $5($\frac{7}{2}$,$11$)$\oplus $($4$,$\frac{21}{2}$)$\oplus $3($4$,$\frac{23}{2}$)$\oplus $($\frac{9}{2}$,$12$)\\ \hline\end{tabular}

   \vskip 10pt
\caption{The GV invariants $n^{d}_{j_L,j_R} = \sum_{d_1+d_2=d} n^{d_1,d_2}_{j_L,j_R} $ for $d=1,2,\cdots, 8$ for the local $\mathbb{P}^1\times \mathbb{P}^1$ model. Here $d_1,d_2$ denote the degrees of the base $\mathbb{P}^1$ and the fiber $\mathbb{P}^1$. There is a symmetry $n^{d_1,d_2}_{j_L,j_R} = n^{d_2,d_1}_{j_L,j_R}$ since the fibration is trivial. }
\label{tableP1P1}

\end{center}
\end{table}

\begin{table}
\begin{center}

 \begin{tabular} {|c|l|} \hline $d$  & $\sum_{d_B+2d_F=d} \sum_{j_L,j_R} \oplus n^{d_B,d_F}_{j_L,j_R}(j_L,j_R)$    \\ \hline 1 & ($0$,$0$)\\ \hline2 & ($0$,$\frac{1}{2}$)\\ \hline3 & ($0$,$1$)\\ \hline4 & \\ \hline5 & ($0$,$2$)\\ \hline6 & ($0$,$\frac{5}{2}$)\\ \hline7 & ($0$,$3$)\\ \hline8 & ($0$,$\frac{5}{2}$)$\oplus $($0$,$\frac{7}{2}$)$\oplus $($\frac{1}{2}$,$4$)\\ \hline9 & ($0$,$3$)$\oplus $($0$,$4$)$\oplus $($\frac{1}{2}$,$\frac{9}{2}$)\\ \hline10 & ($0$,$\frac{5}{2}$)$\oplus $($0$,$\frac{7}{2}$)$\oplus $2($0$,$\frac{9}{2}$)$\oplus $($\frac{1}{2}$,$4$)$\oplus $($\frac{1}{2}$,$5$)$\oplus $($1$,$\frac{11}{2}$)\\ \hline\end{tabular}

    \vskip 10pt
\caption{The GV invariants $n^{d}_{j_L,j_R} = \sum_{d_B+2d_F=d} n^{d_B,d_F}_{j_L,j_R} $ for $d=1,2,\cdots, 10$ for the local $\mathbb{F}_1$ model. Here $d_B,d_F$ denote the degrees of the base $\mathbb{P}^1$ and the fiber $\mathbb{P}^1$.}
\label{tableF1}

\end{center}
\end{table}

\end{document}